\documentclass[12pt]{article}

\usepackage{amsmath,amssymb,amsfonts,booktabs,multirow}
\usepackage[a4paper, left=2.5cm, right=3cm, top=2.5cm, bottom=2.5cm]{geometry}
\usepackage{caption}
\usepackage{graphicx}
\usepackage[numbers]{natbib}
\usepackage{textcomp, setspace}
\usepackage{xcolor,arydshln}
\usepackage{amsthm,supertabular}
\DeclareMathOperator{\sign}{sign}

\usepackage{lmodern}

\graphicspath{{Figs/}}

\newcommand{\ModelName}{JPoNG}

\definecolor{black}{rgb}{0.0, 0.0, 0.0}

\usepackage{lmodern}

\definecolor{electricultramarine}{rgb}{0.25, 0.0, 1.0}
\definecolor{palatinateblue}{rgb}{0.15, 0.23, 0.89}
\definecolor{hanpurple}{rgb}{0.32, 0.09, 0.98}
\definecolor{mediumblue}{rgb}{0.0, 0.0, 0.8}
\newcommand{\bt}[1]{\textcolor{black}{#1}}

\begin{document}

\title{Electric-Gas Infrastructure Planning for Deep Decarbonization of Energy Systems}

\author{Rahman Khorramfar\thanks{Corresponding author. $\hspace{12cm} $ Rahman Khorramfar is affiliated with MIT Energy Initiative (MITEI) and Laboratory for Information $\&$ Decision Systems (LIDS); Dharik Mallapragada is affiliated with MITEI; Saurabh Amin is affiliated with Civil and Environmental Engineering (CEE), MITEI, and LIDS, Massachusetts Institute of Technology, Cambridge, MA, USA. \{khorram, dharik, amins\}@mit.edu.  $\hspace{5cm} $ Authors acknowledge support from  MIT Energy Initiative Future Energy System Center, MIT Climate Grand Challenge ``Preparing for a new world of weather and climate extremes'', and MIT Energy Initiative Project Hurricane Resilient Smart Grids. }, Dharik Mallapragada, Saurabh Amin}

            
\date{}
\maketitle

\begin{abstract}
The transition to a deeply decarbonized energy system requires coordinated planning of infrastructure investments and operations serving multiple end-uses while considering technology and policy-enabled interactions across sectors. Electricity and natural gas (NG), which are vital vectors of today's energy system, are likely to be coupled in different ways in the future, resulting from increasing electrification, adoption of variable renewable energy (VRE) generation in the power sector and policy factors such as cross-sectoral emissions trading. This paper develops a least-cost investment and operations model for joint planning of electricity and NG infrastructures that considers a wide range of available and emerging technology options across the two vectors, including carbon capture and storage (CCS) equipped power generation, low-carbon drop-in fuels (LCDF) as well as long-duration energy storage (LDES). The model incorporates the main operational constraints of both systems and allows each system to operate under different temporal resolutions consistent with their typical scheduling timescales. We  apply our modeling framework to evaluate power-NG system outcomes for the U.S. New England region under different technology, decarbonization goals, and demand scenarios. Under a global emissions constraint, ranging between 80-95\% emissions reduction compared to 1990 levels, the least-cost solution relies significantly on using the available emissions budget to serve non-power NG demand, with power sector using only \bt{14-23\%} of the emissions budget. Increasing electrification of heating in the buildings sector results in greater reliance on wind and NG-fired plants with CCS and results in similar or slightly lower total system costs as compared to the business-as-usual demand scenario with lower electrification of end-uses. Interestingly, although electrification reduces non-power NG demand, it leads to up to \bt{24\%} increase in overall NG consumption (both power and non-power) compared to the business-as-usual scenarios, resulting from the increased role for CCS in the power sector. The availability of low-cost LDES systems reduces the extent of coupling of electricity and NG systems by significantly reducing fuel (both NG and LCDF) consumption  in the power system 
compared to scenarios without LDES, while also reducing total systems costs by up to \bt{4.6\%} for the evaluated set of scenarios. 

\end{abstract}



\section{Introduction}\label{sec:introduction}

Electricity power and NG are important vectors for energy systems in the U.S. and other regions, with the infrastructure for their supply becoming increasingly coupled over the past two decades. For instance, electricity and NG represented 32\% of final energy demand in the U.S. as of 2021 as compared to 22\% in 2005 \cite{US-energy-fact-EIA}. Over the same period, the share of NG-based power generation in the electricity generation mix doubled to 38\% of supply in 2021. This shift away from coal to NG and gradually increasing share of variable renewable energy (VRE) has resulted in the U.S. CO$_2$ emissions being 32\% lower in 2021 compared to 2005 levels \cite{US-energy-today-EIA}. In many U.S. regions with rapidly growing VRE generation, like California and Texas, the flexibility of NG power plants, both within a day and across days, is often used for managing VRE supply and demand variations and ensuring reliable grid operations \cite{KasserisEtal2020}. The importance of NG for reliable power systems operations was exemplified by the Texas freeze of February 2021 where outages in the NG system were identified as the primary cause of electricity blackouts \cite{BusbyEtal2021}. Beyond operations, the planning of electricity and NG infrastructure is becoming more coupled owing to the growing policy interest in electrification for achieving economy-wide decarbonization by mid-century. 

In this article, we develop a model for joint planning of electricity and NG infrastructure that considers spatial, temporal, and technologically resolved representation of the system operations. This model enables us to systematically evaluate the pathways for deep decarbonization of these systems and the other end-use sectors where these vectors are used (e.g. buildings) under plausible technology, demand, and policy scenarios. 

Economy-wide CO$_2$ emissions reduction efforts further increase the inter-dependency between electricity and NG infrastructure planning, beyond levels seen currently. First, electrification of end-uses, such as heating, in the building sector reduces NG consumption in favor of increasing electricity consumption, which can increase peak electricity demand and its timing (e.g. summer to winter peaking shifts \cite{E3Team2020, EFS2021}.) Second, increasing the share of VRE generation in the power sector coupled with the adoption of flexible short-duration and long-duration energy storage resources is likely to alter the cost-optimal operational schedule of NG power plants, and hence impact the pattern of NG withdrawals from the gas infrastructure for power generation. Ultimately, these patterns affect the cost-effectiveness of the available NG-based power generation technologies and the incentive to invest in new NG pipelines. In particular, reduced capacity utilization of NG-power generation is expected to promote more investments in higher operating cost, lower capital cost combustion turbines in comparison to lower operating cost, higher capital cost combined cycle gas turbine (CCGT) plants \cite{ColeEtal2021, JayadevEtal2020}. Third, the scope and stringency of emissions reduction policies can shape the investments in NG and electricity infrastructure in several ways. For example, policies aimed at \textit{combined} emission reductions without specific sectoral emissions reduction requirements may facilitate emissions trading across sectors to minimize the system cost of decarbonization.

\bt{Furthermore, the emerging low-carbon technologies will also impact electric-gas network interactions in distinct ways. These include: }CCGT equipped with carbon capture and storage (CCS) for power generation, long-duration energy storage (LDES) as well as the use of synthetic or renewable-sourced low-carbon drop-in fuels (LCDFs), whose availability and cost is highly uncertain \cite{JayadevEtal2020, LeeEtal2021}. \bt{LCDFs that are compatible with NG infrastructure can be sourced from multiple pathways including conversion of biomass \cite{GassnerFrancois2012, MallapragadaEtal2014}, valorization of waste feedstocks (e.g. municipal waste, manure, fats oils and greases, waster water treatment plants \cite{LeeEtal2021}), as well as production using electricity and renewable hydrogen in conjunction with CO$_2$, captured from the atmosphere \cite{ChoeEtal2021}. Consistent with other energy system modeling studies \cite{ColeEtal2021, HargreavesEtal2020}, 
we assume that LCDF costs substantially more than fossil NG and it is a carbon-neutral fuel in the sense that the combustion emissions associated with its end-use are equal to the atmospheric CO$_2$ emissions captured during its production. These estimates are within the range of life cycle carbon footprint and costs reported for LCDFs compatible with NG infrastructure. For instance, recent life cycle analysis studies suggest that LCDF sourced from waste feedstocks could have negative to slightly positive life cycle GHG emissions \cite{LeeEtal2021}.} Understanding the interplay between these options requires models for coordinated planning of electricity and NG infrastructure that consider investments and operations of generation, transmission, storage, and end-use for each vector, which is the focus of our article.

Traditionally, generation and transmission expansion for electricity systems have been treated as separate problems in the literature \cite{KoltsaklisDagoumas2018}. However, in recent years, there has been a growing interest in integrated generation transmission expansion planning (GTEP)\cite{ZhangEtal2016, DingEtal2017, BodalEtal2020, VonWaldEtal2022, LiEtal2022}. These studies are motivated by the growing importance of VRE generation in power systems and the spatially variable nature of these resources that make transmission expansion more important \cite{BrownBotterud2021, NREL-seams-Interconnections}. Still, the majority of GTEP models primarily focus on power systems without accounting for the operational interdependencies with the NG system. A tri-level model with demand uncertainty is proposed in \cite{PozoEtal2012} in which the upper level is the transmission expansion problem, the intermediate level is the generation expansion problem, and the lower level deals with market operation decisions. Aghaei et al. \cite{Aghaeietal2014} develop a probabilistic model incorporating reliability measures in the objective function. Guerra et al. \cite{GuerraEtal2016} propose a model with demand side control, power reserve, and emission constraints. Uncertainty in peak demand and generation output is considered in \cite{BaringoBaringo2017} and a stochastic adaptive robust optimization model is proposed. Jayadev et al. \cite{JayadevEtal2020} develop a cost-optimal, multi-stage planning model for the US power system and show that substantial decarbonization is moderately costly and is accompanied by investments in new NG generation capacity that has declining utilization rates.  Li et al. \cite{LiEtal2022} compare three mixed-integer formulations and develop two Benders algorithms where the first algorithm decomposes the problem based on the planning years, whereas the second one is based on operational decisions. The literature of expansion models in energy systems is rich and a thorough treatment of the subject can be found in review papers such as Dagoumas et al. \cite{DagoumasEtal2019} and Farrokhifar et al. \cite{FarrokhifarEtal2020}. 

Among the few papers studying power and NG infrastructure interactions, Unsihuay et al.  \cite{UnsihuayEtal2010} propose a multi-period mixed-integer linear program (MILP) for the Brazilian integrated power-NG systems \bt{with detailed modeling of NG system supply chain including NG production, storage, and transportation. The results show the importance of joint planning and highlight the complementarities between NG storage and hydropower to hedge against the uncertainty of water inflow to reservoirs. Chaudry et al. \citep{ChaudryEtal2014} consider CCS technologies and develop long-term planning of a power-gas system in Great Britain to achieve a low-carbon energy system. } 
Qiu et al. \cite{QiuEtal2015a} present an AC power flow model \bt{to determine the optimal expansion of gas-fired power plants} and propose linearization methods for the resulting nonlinear program.
Shao et al. \cite{ShaoEtal2017}  consider a joint power and NG transportation system and \bt{develop an adaptive robust optimization problem to enhance power grid resiliency under extreme conditions by utilizing the existing NG infrastructure and replacing overhead transmission lines with underground ones.} Two-stage and multistage stochastic programming approaches are respectively proposed in \cite{ZhaoEtal2017} and \cite{DingEtal2017} for the expansion of gas-fired plants and NG system.
Zhang et al.\cite{zhangEtal2018} propose a security-constrained model that considers $N-1$ contingency in both power and NG system. More recently, Von Wald et al. \cite{VonWaldEtal2022} develop a multi-period generation expansion model with detailed operational constraints. They consider endogenous electrification of the end-use appliances under system-specific emission constraints as well as the role for hydrogen blending and its application in an abstract case study. The model considers both short-duration battery storage and long-duration power storage realized by storing hydrogen obtained by electrolysis. The model, however, does not consider the expansion of power or NG system networks.

\bt{The current literature of power-gas GTEP addresses several important aspects of the joint system including CCS technologies \citep{ChaudryEtal2014, VonWaldEtal2022}, battery storage \citep{VonWaldEtal2022},  NG storage facilities\citep{UnsihuayEtal2010, ChaudryEtal2014}, hydrogen for long-duration power storage \citep{VonWaldEtal2022}, emissions reduction targets \citep{ChaudryEtal2014, VonWaldEtal2022}, and electrification of end-use \citep{VonWaldEtal2022}. However, to our knowledge, no thorough treatment of the role of emerging supply-side technologies such as CCS-equipped power generation, LCDF, and LDES with regard to system outcome and the tradeoff between emissions reduction and build system cost has been undertaken. Moreover, the impact of electrification on NG utilization in power and non-power sectors has received limited attention. In that regard, we model interactions between NG and electricity networks in deeply decarbonized regional-scale energy systems, considering deployment of these complementary technologies to facilitate renewable penetration and demand-side interventions in the form of increased electrification of the building sector. The model fidelity includes representation of both NG and electricity networks as well as hourly and daily supply-demand balancing of the electricity and NG system operations, respectively. Below we summarize the salient features of the proposed modeling framework:}
\begin{itemize}
    \item \bt{We account for the fact that each network 
operates on a different time resolution. This is in contrast to the previous studies where different time resolutions are treated in separate steps \citep{SaediEtal2021}. 
    \item We consider CO$_2$ capture and storage infrastructure investments to ensure that power nodes with CCGT-CCS plants transfer the locally captured CO$_2$ to remote storage locations via dedicated CO$_2$ pipelines.
    \item Our model distinguishes between NG nodes and gas storage nodes which consist of storage, liquefaction and vaporization facilities. 
    \item We analyze the impact of emissions reduction constraint that can be imposed to limit either sectoral or economy-wide emissions. The latter constraint allows emission trading between power and NG systems. We systematically  compare the relative impact of sectoral vs. economy-wide emissions reduction constraints on the planning outcomes.
    \item We consider resource availability constraints that limit the development of VRE sources based on potential land availability.} 
\end{itemize}



We apply our model to a case study of the U.S. New England region where we explore system outcomes for alternative technology, sectoral coordination, demand and deep decarbonization scenarios.\footnote{Our model relies on renewable resource profiles, load projections, technology cost assumptions and approximate gas and electric network topology based on publicly available data (see the Appendix) for the region. Note that the model is representative but not field-validated; hence the results should be interpreted in the context of technology adoption trends and requirements, and not as exact projections for the region.  
} Our results reveal that electrification of the building sector in conjunction with other emerging supply-side mitigation options provides a cost-effective means for steep emission reductions (80-95\% compared to 1990 levels) across both systems and end-uses relying on these vectors. \bt{The results further highlight that although there is an obvious need to deploy substantial quantities of VRE and Li-ion storage to support power system decarbonization, the uncertainty in supply-side and demand-side drivers across infrastructures for electricity and NG leaves open the possibility of multiple pathways for deep decarbonization.}
The rest of the paper is organized as follows. Section~\ref{sec:model-formulation} formulates the joint power-NG system planning model. The details of the case study on the New England region are explained in Section~\ref{sec:case-study}. Results and analysis with key takeaways are discussed in Section~\ref{sec:results}. The concluding remarks and areas of future work are discussed in Section~\ref{sec:conclusion}. 

\section{Model Formulation}\label{sec:model-formulation}
Our model for joint power and NG planning, referred as \ModelName, determines the minimum cost planning decisions for power and NG systems considering the two systems' interdependency. The proposed model considers a range of generation and storage technologies whose operations are modeled via operational and policy constraints across a set of representative periods. The formulation allows different temporal resolutions for the operation of both systems since the data availability or planning requirements can be different for each system. For example, decisions related to power generation, such as dispatch amounts and unit commitment, require hourly resolution. 
NG system operation, however, does not involve generation decisions and only deals with transmission and storage operations for which daily resolution may be sufficient. 
Moreover, due to the ability of NG pipelines to also provide some storage via line packing, daily resolution for scheduling NG operations could facilitate the management of intra-day variations in NG demand. 

In the model, the operations of both systems are coupled through two sets of constraints. The first set ensures NG flow to the power system. The second coupling constraint limits the CO$_2$ emission incurred by consuming fossil-derived NG in both power and NG systems. This is in contrast with previous studies where the emission limit is either applied to power system only \cite{LiEtal2022}, or is imposed separately for each system \cite{VonWaldEtal2022}.

As a common practice in the literate to enhance the tractability of the GTEP \cite{VonWaldEtal2022, LiEtal2022, TeichgraeberBrandt2022}, the scheduling of power system operations is modeled over a set of representative days. 
Long-term planning of NG system is usually planned on yearly \cite{QiuEtal2015a,BaratiEtal2014,SaldarriagaEtal2019}, monthly \cite{HamediEtal2009}, or daily \cite{ChaudryEtal2014,VonWaldEtal2022} basis. 
Some studies use different resolutions for power and NG systems. Power and gas systems operate on an hourly basis and daily basis over representative time periods in Von Wald et al. \cite{VonWaldEtal2022}. Saedi et al. \citep{SaediEtal2021} used half-hour resolution for the power system and daily resolution for the NG. The authors propose a  two-phase model in which the power system operations are planned in the first phase and the second phase determines the optimal gas flow. We take a similar time resolution here and consider hourly and daily time resolution for the power and NG systems, respectively. However, we only consider representative days for the power system and model the operations of the NG system across all days of the year. 

\begin{figure*}
    \centering
    \includegraphics[width=0.8\textwidth]{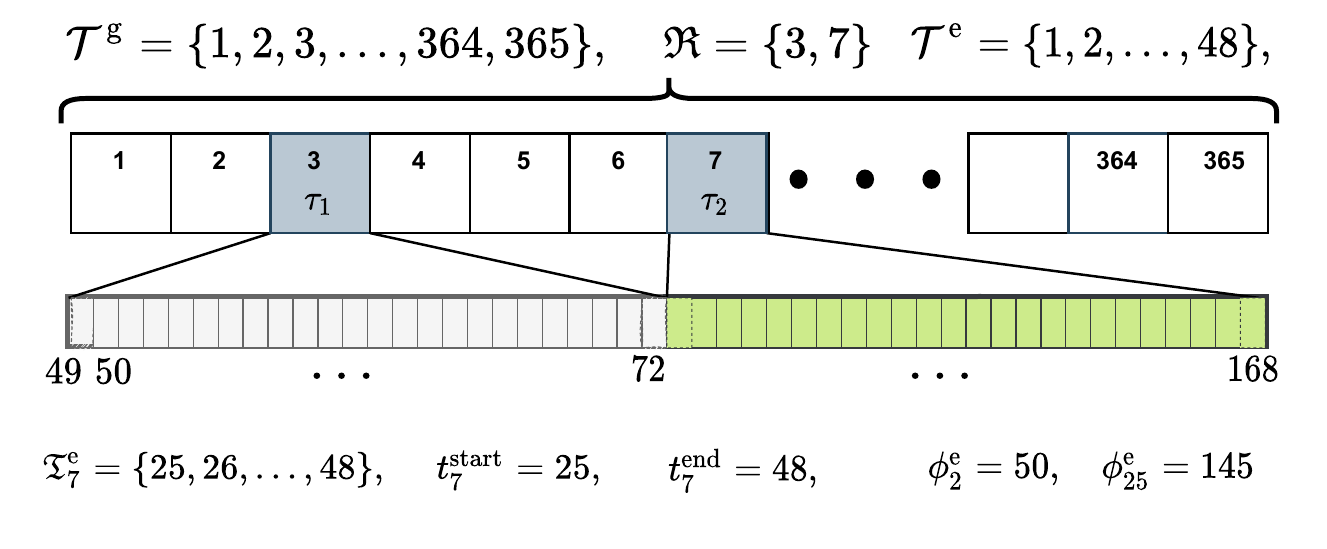}
    \caption{Illustration of the model's temporal resolution  for the case of two representative days used for power systems operations. The top row represents planning days in a year, with days 3 and 7 as representative days forming the set $\mathfrak{R}$. The bottom row shows the hours corresponding to representative days. The set of planning periods for the NG system ($\mathcal{T}^{\text{g}}$) is the entire year. The set of planning periods for the power system ($\mathcal{T}^{\text{e}}$) is the chronologically ordered set of hours in the representative days whose mapping to their original index is given by $\phi^{\text{e}}_t, t\in \mathcal{T}^{\text{e}}$. The set of hours in their original indexing is denoted by $\mathfrak{T}^{\text{e}}_\tau$ with $t^{\text{start}}_\tau$ and $t^{\text{end}}_\tau$ signifying the starting and ending hours for the representative day $\tau$.  
    }
    \label{fig:planning-res}
\end{figure*}

Although the sub-daily representation can be more advantageous in capturing fine-grained operational features, daily planning in the context of capacity expansion has several merits. First, it enhances the scalability of the model by significantly reducing the number of decision variables and constraints, which allows for introducing complexity in other dimensions of the problem, such as using integer variables for investment decisions, as is done in this paper. Second, it no longer necessitates using nonlinear steady-state gas flow that relates nodal pressure to flow. In other words, a non-compressible formulation can be employed for gas flow in case of daily resolution in which the gas is treated as liquid \cite{PabloThesis2013}. The third advantage stems from the data availability. In many cases, data to characterize the NG network operations for real-world systems, such as our case study, is not publicly available at the resolution needed for hourly operations considering pressure variations. 
Fig.~\ref{fig:planning-res} illustrates our modeling approach for the time resolutions of both systems.

\begin{figure}
    \centering
    \includegraphics[width=0.4\textwidth]{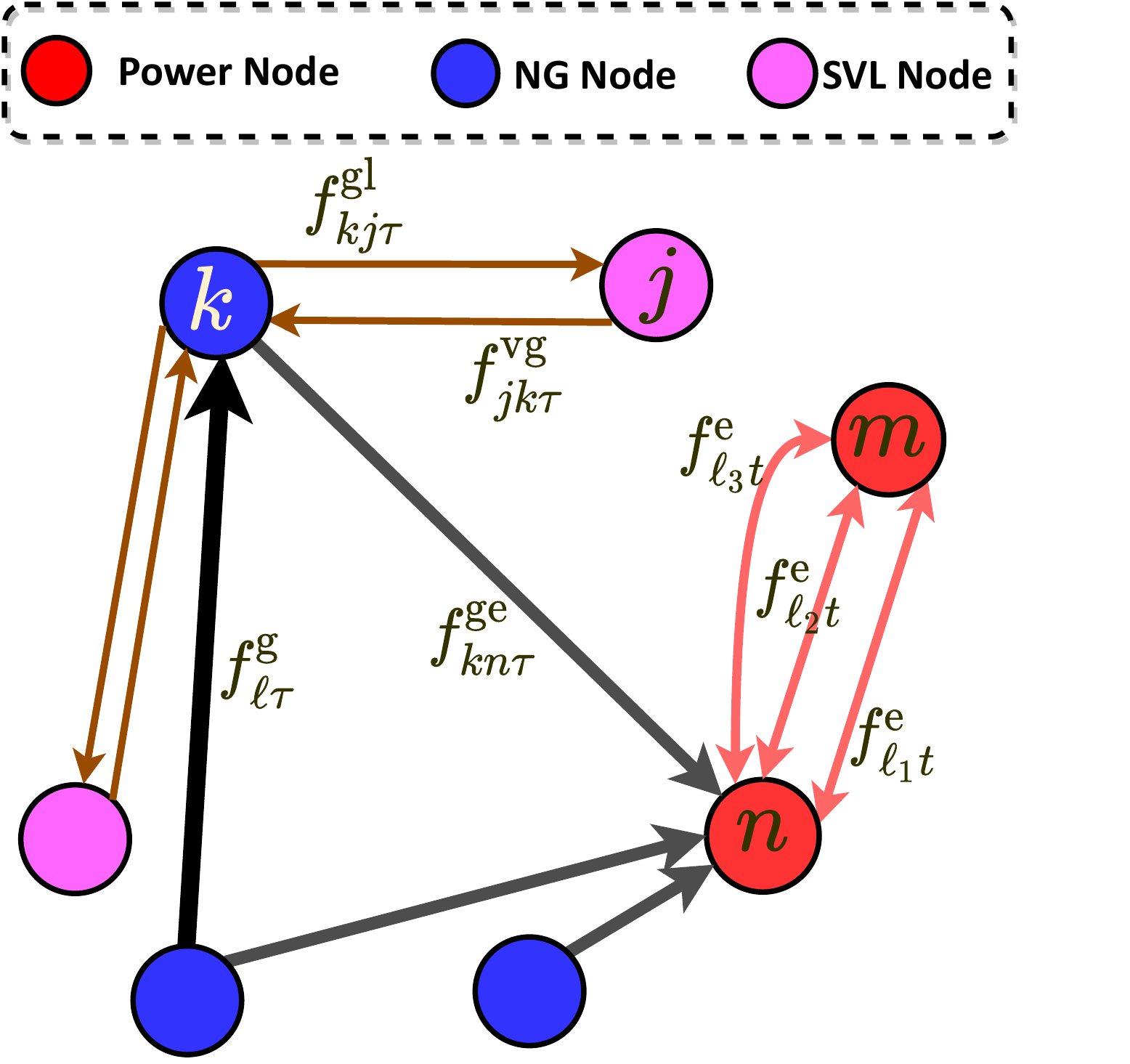}
    
    \caption{Flow variables between different nodes. In our model, power nodes can be connected by multiple bi-directional transmission lines denoted by $f^{\text{e}}_{\ell t}$.  Each power node operates a set of local gas-fired plants by drawing gas from nodes that are connected to it. The variable $f^{\text{ge}}_{kn \tau}$ captures this flow. 
    Each NG node is connected to its adjacent SVL nodes through two unidirectional pipelines where one is from NG to SVL's liquefaction facilities denoted by $f^{\text{gl}}_{kj \tau}$; and the other one from SVL's vaporization facility to NG node denoted by $f^{\text{vg}}_{jk\tau}$. The variable $f^{\text{g}}_{\ell \tau}$ denotes the pipe flow between NG nodes. NG nodes can be connected by one or more uni-directional pipelines, but only one connection is depicted here. Candidate transmission lines and pipelines are not shown in this figure.}
    \label{fig:node-connections}
\end{figure}

The network representation in the model consists of three sets of nodes as depicted in Figure~\ref{fig:node-connections}. The first set represents power system nodes and is characterized by different generation technologies (i.e., plant types), demand, storage, and the set of adjacent nodes by which the node can exchange electricity. The second set of nodes are NG nodes, each associated with injection amount, demand, and its adjacent nodes. Storage tanks, vaporization and liquefaction facilities, which  are commonly used in the non-reservoir storage of NG, collectively form the third set of nodes referred to as  storage-vaporization-liquefaction (SVL) nodes. We allow for the possibility of NG storage infrastructure to be located far from demand or injection points in the network, as per existing practice \cite{NGstatGuide2021} (see \ref{app-NG-topology} for detailed discussion). Accordingly, our model makes a distinction between NG and SVL nodes to account for their distinct locations. 
We also allow the NG system to use LCDF which represents a renewable source of fuel that is interchangeable with NG and hence can be imported and transported by the NG pipelines \cite{ColeEtal2021}.

\bt{The objective function of \ModelName{} is the sum total of investment and operational costs for power and NG system. The constraints of \ModelName{} are the set of all constraints for power and NG systems as well as constraints that couple the operations of both systems together.} We present the power and NG systems' model as well as coupling constraints separately for ease of exposition. The full description of the mathematical notation used in the formulation is described in \ref{app:nomenclature}.

\subsection{Power System Model}
\noindent\textbf{Objective Function:}
\begin{subequations}
\label{elec-obj}
\begin{align}
      \min & \sum_{n \in \mathcal{N}^{\text{e}}} \sum_{i \in \mathcal{P}}  (C^{\text{inv}}_i x^{\text{est}}_{ni}+C^{\text{fix}}_{i} x^{\text{op}}_{ni} + \sum_{r \in \mathcal{S}^{\text{e}}_n}(C^{\text{pInv}}_r+C^{\text{pFix}}) y^{\text{eCD}}_{nr})\notag\\
       &+\sum_{n \in \mathcal{N}^{\text{e}}} \sum_{r \in \mathcal{S}^{\text{e}}_n}(C^{\text{EnInv}}_r+C^{\text{EnFix}}) y^{\text{eLev}}_{nr}   \label{elec-obj-1}\\
        & +\sum_{n \mathcal{N}^{\text{e}}} \sum_{i \in \mathcal{P}}  C^{\text{dec}}_i x^{\text{dec}}_{ni} \label{elec-obj-2}\\
    &  +\sum_{n \in \mathcal{N}^{\text{e}}}\sum_{i \in \mathcal{P}} \sum_{t \in \mathcal{T}^{\text{e}}}w_t p_{nti}C^{\text{var}}_{i}  \label{elec-obj-3}\\
   &+\sum_{n \in \mathcal{N}^{\text{e}}}\sum_{t \in \mathcal{T}^{\text{e}}}\sum_{i \in \mathcal{P}} w_t C^{\text{startUp}}_i  x^{\text{up}}_{nti} \label{elec-obj-4}\\
      &+ \sum_{l\in \mathcal{L}^{\text{e}}}  C^{\text{trans}}_{\ell} z_{\ell}^{\text{e}} 
   \label{elec-obj-5} \\ 
      &+\sum_{n\in \mathcal{N}^{\text{e}}} d_n C^{\text{inv}}_{\text{CO}_2}\kappa^{\text{pipe}}_{n}+ C^{\text{str}}_{\text{CO}_2} \sum_{n\in \mathcal{N}^{\text{e}}} \sum_{t\in \mathcal{T}^{\text{e}}} w_t \kappa^{\text{capt}}_{nt}\label{elec-obj-6}\\ 
         &+  \sum_{n \in \mathcal{N}^{\text{e}}}\sum_{i \in \mathcal{P}} \sum_{t \in \mathcal{T}^{\text{e}}}w_t p_{nti} (C^{\text{fuel}}_{i} h_i)  \label{elec-obj-7}\\
    &+\sum_{n \in \mathcal{N}^{\text{e}}} \sum_{t \in \mathcal{T}^{\text{e}}}w_t C^{\text{eShed}}_n a^{\text{e}}_{nt}
    \label{elec-obj-8}
\end{align}
\end{subequations}

The objective function~\eqref{elec-obj} minimizes the  total investment and operating costs incurred in the power system. The first term~\eqref{elec-obj-1} is the investment and fixed operation and maintenance (FOM) costs for generation and storage. The term~\eqref{elec-obj-2} captures the cost of plant retirement or decommissioning. The variable operating and maintenance  (VOM) cost is represented by the term~\eqref{elec-obj-3}. The fourth term~\eqref{elec-obj-4} corresponds to  startup costs of power plants with unit commitment constraints (i.e. thermal power plants). The network expansion costs is included in the term~\eqref{elec-obj-5}. The cost of CO$_2$ transport and storage infrastructure required to accompany CCGT-CCS power generation is incorporated by term~\eqref{elec-obj-6} which also captures the cost associated with establishing CO$_2$ pipelines and storage. Here, we conservatively assume that each CO$_2$ pipeline connects a power node to the storage site, which ignores the possibility of meshed network design for CO$_2$ transport. The cost of fuel consumption for non NG-fired power plants (i.e., nuclear) is ensured by term~\eqref{elec-obj-7}.  The term~\eqref{elec-obj-8} penalizes the load shedding in the power system. 

\noindent\textbf{Investment and Unit Commitment:} For every $n\in \mathcal{N}^{\text{e}},i\in \mathcal{P}$
\begin{subequations}
\begin{align}
      & x^{\text{op}}_{ni} = I^{\text{num}}_{ni}-x^{\text{dec}}_{ni}+x^{\text{est}}_{ni} &  \label{elec-c1}\\
   &x_{nti} - x_{n,t-1,i} = x^{\text{up}}_{nti}-x^{\text{down}}_{nti}& t\in \mathcal{T}^{\text{e}} \label{elec-c2}\\
    &x_{nti} \leq x^{\text{op}}_{ni}&t\in \mathcal{T}^{\text{e}} \label{elec-c3}
\end{align}
\end{subequations}
Constraints~\eqref{elec-c1} model the number of operating plants.  The unit commitment constraints for each node and plant type are presented in \eqref{elec-c2} which computes the number of plants committed, started up, or shut down across the hours of the representative days. Constraints~\eqref{elec-c3} limit the number of committed units to the number of available ones. As per other similar studies \cite{PalmintierWebster2015,SepulvedaEtal2021,LiEtal2022,VonWaldEtal2022}, we relaxed the integrality of unit commitment decisions which has been shown to introduce relatively small numerical error while significantly reducing the computational complexity of the problem.

\noindent\textbf{Generation, Ramping, and Load Shedding:}
For every $n \in \mathcal{N}^{\text{e}},t\in \mathcal{T}^{\text{e}}$
\begin{subequations}
\begin{align}
    &L^{\text{prod}}_{i} x_{nti} \leq  p_{nti} \leq   U^{\text{prod}}_{i}x_{nti}&\hspace{-2cm} i\in \mathcal{H} 
    \label{elec-c4}\\
    &|p_{nti} -p_{n,(t-1),i}| \leq   U^{\text{ramp}}_{i} U^{\text{prod}}_{i} (x_{nti}-x^{\text{up}}_{nti})+\notag\\
    &\max(L^{\text{prod}}_{i}, U^{\text{ramp}}_{i})U^{\text{prod}}_{i}x^{\text{up}}_{nti}   & \hspace{-2cm} i\in \mathcal{H} 
    \label{elec-c5}\\
     &p_{nti} \leq \rho_{nti}U^{\text{prod}}_{i} x^{\text{op}}_{ni} &\hspace{-2cm} i \in \mathcal{R}  \label{elec-c6}\\
     & a^{\text{e}}_{nt}\leq D^{\text{e}}_{n\phi^{\text{e}}_t}& \label{elec-c7} 
\end{align}
\end{subequations}

The generation limits are imposed in constraints~\eqref{elec-c4}. 
Constraints~\eqref{elec-c5} are the ramping constraints that limit the  generation difference of thermal units in any consecutive time periods to a ramping limit in the right-hand-side of the equation. The generation pattern of VREs is determined by their hourly profile in the form of capacity factor; constraints~\eqref{elec-c6} limit the generation of VRE to hourly capacity factor (i.e. $\rho_{nti}$) of maximum available capacity (i.e. $U^{\text{prod}}_{i}x^{\text{op}}_{ni}$). Constraints~\eqref{elec-c7} state that the load shedding amount can not exceed demand. Note that we use the mapping $\phi^{\text{e}}_t$ to access the demand in the corresponding hour of a representative day.

\noindent\textbf{Power Balance Constraints:}
For every $n\in \mathcal{N}^{\text{e}}, t\in \mathcal{T}^{\text{e}}$

\begin{align}
        &\sum_{i \in \mathcal{P}}p_{nti} +\sum_{m\in \mathcal{N}^{\text{e}}}\sum_{l\in \mathcal{L}^{\text{e}}_{nm}}\sign(n-m) f^{\text{e}}_{\ell t}+\notag
        \sum_{r\in \mathcal{S}^{\text{e}}_n} (s^{\text{eDis}}_{ntr}-s^{\text{eCh}}_{ntr})  \\ 
        &+ a^{\text{e}}_{nt}=D^{\text{e}}_{n \phi^{\text{e}}_t}+d_n E^{\text{pipe}}\kappa^{\text{pipe}}_{n}+ E^{\text{cprs}} E^{\text{pump}}\kappa^{\text{capt}}_{nt}
    & \hspace{-2cm} 
    \label{elec-c8}
\end{align}

Constraints~\eqref{elec-c8} ensure that for each node and for each planning period the generation, the net flow, the net storage, and the load shedding amount should be equal to the net demand. The net demand is defined in the right-hand-side where the first term is the baseline demand, the second term is the electricity consumption by CO$_2$ pipelines and the last term is the electricity used by compressors.  
The notation $\sign(n-m)$ is the \textit{sign} function that takes value -1 if $n<m$, value 1 if $n>m$, and 0 otherwise.
We use this function to ensure that $f^e_{\ell t}$ appears with opposite signs (i.e., negative of positive signs) in the balance equations of the nodes connected by transmission line $\ell$. 

\noindent\textbf{Network Constraints:} 
For every $l\in \mathcal{L}^{\text{e}},t \in \mathcal{T}^{\text{e}}, \text{ and } n,m\in \mathcal{N}^{\text{e}}_\ell$

\begin{subequations}
\begin{align}
    & |f^{\text{e}}_{lt }| \leq I^{\text{trans}}_{\ell}&\hspace{-2cm}  \text{if } \mathcal{I}^{\text{trans}}_\ell=1  
    \label{elec-c9}\\
      & |f^{\text{e}}_{lt }| \leq U^{\text{trans}}_\ell z^{\text{e}}_{\ell}& \hspace{-2cm}  \text{if } \mathcal{I}^{\text{trans}}_\ell=0  \label{elec-c10}\\
&|f^{\text{e}}_{\ell t} - b_{\ell}(\theta_{mt}-\theta_{nt})|\leq M(1-z^{\text{e}}_{\ell}) & \text{if } \mathcal{I}^{\text{trans}}_\ell=0 
\label{elec-c11}\\
        &f^{\text{e}}_{\ell t} = b_{\ell}(\theta_{mt}-\theta_{nt})&\hspace{-2cm} \text{if } \mathcal{I}^{\text{trans}}_\ell=1  \label{elec-c12}\\
    & \theta_{0,t}=0  &  \label{elec-c14}
\end{align}
\end{subequations}
Flow for the existing transmission lines is limited by constraints~\eqref{elec-c9}. Constraints~\eqref{elec-c10} limit the flow in candidate transmission lines only if it is already established (i.e., $z^{\text{t}}_{\ell}$=1). Throughout the paper, we use $M$ to denote a big number. 
Direct Current (DC) power flow constraints for candidate and existing transmission lines are respectively imposed in constraints~\eqref{elec-c11} and \eqref{elec-c12}. The phase angle for the reference node $0$ is set in constraint~\eqref{elec-c14}.

\noindent\textbf{Storage Constraints:} For every $n\in \mathcal{N}^{\text{e}}, r\in \mathcal{S}^{\text{e}}_n$
\begin{subequations}
\begin{align}
 s^{\text{eLev}}_{n t^{\text{start}}_\tau r} &=(1-\gamma^{\text{loss}}_{r}) (s^{\text{eLev}}_{n,t^{\text{end}}_\tau r}-s^{\text{rem}}_{n\tau r})+\notag \\
 & \gamma^{\text{eCh}}_r s^{\text{eCh}}_{nt^{\text{start}}_\tau r}-\frac{s^{\text{eDis}}_{n t^{\text{start}}_\tau r}}{\gamma^{\text{eDis}}_r},\qquad  \tau \in \mathfrak{R} & \label{elec-c15}\\
  s^{\text{eLev}}_{n t r} &=(1-\gamma^{\text{loss}}_{r}) (s^{\text{eLev}}_{ntr})+ \gamma^{\text{eCh}}_r s^{\text{eCh}}_{ntr}-\frac{s^{\text{eDis}}_{n t r}}{\gamma^{\text{eDis}}_r}&\notag \\
  & \hspace{3cm} t\in \mathcal{T}^{\text{e}}\backslash \{t^{\text{start}}_\tau \lvert \  \tau \in \mathfrak{R} \}   \label{elec-c16}\\
  s^{\text{day}}_{n,\tau+1,r} & = (1-24\gamma^{\text{loss}}_{r})s^{\text{day}}_{n,\tau,r}+s^{\text{rem}}_{n\Omega_{\tau},r}, \quad \tau \in \mathcal{T}^{\text{g}}\backslash 365 \label{elec-c17}\\
s^{\text{day}}_{n,1,r} & = (1-24\gamma^{\text{loss}}_{r})s^{\text{day}}_{n,\tau,r}+s^{\text{rem}}_{n\Omega_{\tau},r},\quad \tau = 365& \label{elec-c18}\\
s^{\text{day}}_{n\tau r} & = s^{\text{eLev}}_{nt^{\text{end}}_\tau r}-s^{\text{rem}}_{n\tau r}, \quad \tau \in \mathfrak{R} \label{elec-c19}\\
s^{\text{rem}}_{n\tau r}&=0, \quad \tau \in \mathfrak{R}, r\in \mathcal{S}^{\text{eS}} \label{elec-c20}\\
    & s^{\text{eCh}}_{ntr}\leq y^{\text{eCD}}_{nr}  &  \label{elec-c21}\\
    &s^{\text{eCh}}_{ntr}\leq y^{\text{eCD}}_{nr}  &  \label{elec-c21-2}\\
    & s^{\text{eLev}}_{ntr}\leq y^{\text{eLev}}_{nr} &  \label{elec-c22}
\end{align}
\end{subequations}

Recall (see Fig.~\ref{fig:planning-res}) that representative days are not necessarily consecutive which needs further constraints to account for the potential \textit{carryover} of energy between representative days, which is particularly important when modeling LDES. Li et al. \cite{LiEtal2022} enforce the beginning and ending storage levels of each representative day to 50\% of the maximum storage level. Here, we use a similar approach for short-duration batteries in which we time-wrap the beginning and ending hours of a day. In particular, we assume the same state of charge for storage for the beginning and ending hours of a day for short-duration storage, implicitly precluding energy carryover between representative days. 

For LDES, however, we use the method proposed in \cite{GenX2017, KotzurEtal2018} in which the unrestricted variable $s^{\text{rem}}_{n\tau r}$ models the energy carryover across all calendar days. \bt{This formulation allows LDES to store energy for periods longer than a day (e.g. multiple days), which may be attractive given its lower energy capital cost compared to short-duration storage  like Li-ion. The variables $s^{\text{day}}_{n\tau,r}$ and $s^{\text{rem}}_{n\omega_\tau,r}$ are respectively defined for all days of the planning period (i.e., one year) and representative days where the former captures the storage level at beginning of days, and the latter is the storage carryover during representative days. }

Constraints~\eqref{elec-c15} model battery storage dynamics for the initial hours of each representative day. \bt{ The first term in the right-hand-side is the storage level at the end of the day minus the amount of energy carried over to the next representative day, both adjusted with the loss factor. The second and third terms are effective charge and discharge amounts.}
Constraints~\eqref{elec-c16} model the storage balance for the remaining hours.  
For each day of the planning horizon, the energy transfer across consecutive days is modeled via constraints~\eqref{elec-c17}. The storage in the first and last calendar days is related by constraints~\eqref{elec-c18}. Constraint~\eqref{elec-c19} only applies to representative days and ensures that the storage at the beginning of a day is equal to the storage level at the last hour of the day minus the storage carryover. The storage carryover for short-duration batteries is prevented by constraints~\eqref{elec-c20}, \bt{- in other words, only LDES technologies allowed to carry over energy across calendar days}. Finally, the charge/discharge limits on storage level are imposed in constraints~\eqref{elec-c21}  to \eqref{elec-c22}. Analogous to similar studies on power system expansion \cite{SepulvedaEtal2021, LiEtal2022}, we do not account for use-dependent storage capacity degradation.

\noindent\textbf{Renewable Portfolio Standards (RPS):} 

\begin{align}
   & \sum_{n\in \mathcal{N}^{\text{e}}}\sum_{t\in \mathcal{T}^{\text{e}}}\sum_{i\in \mathcal{R}}  p_{nti} \geq L^{\text{RPS}}\sum_{n\in \mathcal{N}^{\text{e}}}\sum_{t\in \mathcal{T}^{\text{e}}} D^{\text{e}}_{n\phi^{\text{e}}_t}  
   \label{elec-c23}
\end{align}

RPS is a policy constraint that requires a minimum share of the total electricity to be met by renewable sources and has been one of the core policies driving decarbonization efforts in many U.S. and global regions \cite{E3Team2020, MITEI2022FES}. In the model, constraint~\eqref{elec-c23} requires that annual dispatched renewable generation must be at least a pre-specified fraction of total annual demand, specified by the parameter, $L^{RPS}$.

\noindent\textbf{Resource Availability Constraints:} 
\begin{align}
   & \sum_{n\in \mathcal{N}^{\text{e}}}\sum_{i\in \mathcal{Q}}  U^{\text{prod}}_i x^{\text{op}}_{ni} \leq U^{\text{prod}}_{\mathcal{Q}} & \mathcal{Q}\in \mathcal{Q}'
   \label{elec-c24}
\end{align}

We consider resource availability limits for the development of VRE sources. In comparison to thermal plants, the siting of renewable resources is a major challenge due to the relatively large land area footprint per MW, the spatial heterogeneity in their resource availability and land availability limits due to non-energy considerations such as preserving the natural landscape \cite{E3Team2020}. Therefore, constraint~\eqref{elec-c24} limits the installed capacity of a certain set of power plants to their maximum availability limit. The parameter $\mathcal{Q}$ denotes a generation technology class for which there is a resource availability limit. These classes include solar, onshore wind, offshore wind, and nuclear and are represented by set $\mathcal{Q}'$. Note that each technology class can include multiple plant types. For example, nuclear technology can include existing and new nuclear plant types.  

\noindent\textbf{Carbon Capture and Storage (CCS) Constraints:} 
For every $n\in  \mathcal{N}^{\text{e}}, t\in \mathcal{T}^{\text{e}}$ 
\begin{subequations}
\begin{align}
   &\kappa^{\text{capt}}_{nt} = \eta^{\text{g}} \eta_i h_ip_{nti} & i\in \mathcal{CCS} \label{elec-c25}\\
   &\kappa^{\text{capt}}_{nt}\leq \kappa^{\text{pipe}}_{n} \label{elec-c26}\\
   &\sum_{n\in \mathcal{N}^{\text{e}}}\sum_{t\in \mathcal{T}^{\text{e}}} \kappa^{\text{capt}}_{nt} \leq U^{\text{CCS}}& \label{elec-c27}
\end{align}
\end{subequations}

The constraint~\eqref{elec-c25} computes the amount of captured carbon in gas-fired power plants equipped with CCS technology. Constraint~\eqref{elec-c26} determines the CO$_2$ pipeline capacity. Finally, constraint~\eqref{elec-c27} limits the total amount of captured CO$_2$ to the annual CO$_2$ storage capacity.

\subsection{NG System Model}
We now model the objective function and constraints pertaining to the NG system in the \ModelName.

\noindent\textbf{Objective Function:}
\begin{subequations}\label{ng-obj}
\begin{align}
    \min \ & \sum_{l\in \mathcal{L}^{\text{g}}} C^{\text{pipe}}_{\ell} z_{\ell}^{\text{g}}   \label{ng-obj-1}\\
    &+ \sum_{k \in \mathcal{N}^{\text{g}}} \sum_{\tau \in \mathcal{T}^{\text{g}}}  C^{\text{ng}} g_{k\tau}\label{ng-obj-2}\\
    &+ \sum_{j \in \mathcal{N}^s}(C^{\text{strInv}}_j x^{\text{gStr}}_{j} +C^{\text{vprInv}}_j x^{\text{vpr}}_{j}) \label{ng-obj-3}\\
    &+ \sum_{j \in \mathcal{N}^s}\left(C^{\text{strFix}}(I^{\text{gStr}}_{j}+x^{\text{gStr}}_j)+C^{\text{vprFix}}(I^{\text{vpr}}_{j}+x^{\text{vpr}}_j) \right)\label{ng-obj-4}\\
    &+ \sum_{k \in \mathcal{N}^{\text{g}}} \sum_{\tau \in \mathcal{T}^{\text{g}}} (C^{\text{LCDF}}a^{\text{LCDF}}_{k\tau } +C^{\text{gShed}} a^{\text{ng}}_{k\tau })\label{ng-obj-5}
\end{align}
\end{subequations}

The objective function~\eqref{ng-obj} minimizes the total investment and operating costs incurred in the NG system. The first term~\eqref{ng-obj-1} is the investment cost for establishing new pipelines. The second term~\eqref{ng-obj-2} is the cost of procuring NG from various sources to the system. For example, New England procures its NG from Canada and its adjacent states such as New York. 
The term~\eqref{ng-obj-3} and ~\eqref{ng-obj-4} handle the investment and FOM costs associated with NG storage, respectively. The last term~\eqref{ng-obj-5} captures the cost of using LCDF and NG load shedding.  

\noindent\textbf{NG Balance Constraint:}  For every $k\in \mathcal{N}^{\text{g}}, \tau \in \mathcal{T}^{\text{g}} $
\begin{align}
   &g_{k\tau} -\sum_{l \in \mathcal{L}^{\text{gExp}}_{k}} f^{\text{g}}_{\ell\tau}+\sum_{l \in \mathcal{L}^{\text{gImp}}_{k}} f^{\text{g}}_{\ell\tau}-\sum_{n\in \mathcal{A}^{\text{e}}_k} f^{\text{ge}}_{kn\tau } \notag\\
   &+\sum_{j\in \mathcal{A}^s_k} (f^{\text{vg}}_{jk \tau}-f^{\text{gl}}_{kj\tau})+ a^{\text{LCDF}}_{k\tau}+a^{\text{g}}_{k\tau}=D^{\text{g}}_{k\tau} &  \label{ng-c1}
\end{align}

Constraints~\eqref{ng-c1} state that for each node and period, the imported NG (i.e., injection), flow to other NG nodes, flow to power nodes, flow to and from storage nodes, load satisfied by LCDF, and unsatisfied NG load should add up to demand. Unlike power flow, the flow in pipelines is assumed to be unidirectional as it is typical for most long-distance transmission pipelines involving booster compressor stations \cite{VonWaldEtal2022}. We are ignoring the relatively small electricity demand associated with booster compression stations along the NG pipeline network.  

\noindent\textbf{Flow on Representative Days:}

\begin{align}
   &f^{\text{ge}}_{kn\tau_1 } = f^{\text{ge}}_{kn\tau_2 }& \tau_1, \tau_2 \in \mathcal{T}^{\text{g}} \text{ if } \Omega_{\tau_1} = \Omega_{\tau_2} \label{ng-c2-1}
\end{align}

Given the set of representative days used to model power system operations (see Fig.\ref{fig:planning-res}), constraint~\eqref{ng-c2-1} ensures that gas consumption by the power system for all the days represented by the same day is identical.

\noindent\textbf{Gas and LCDF Supply Constraints:}
For every $k\in \mathcal{N}^{\text{g}}, \tau \in \mathcal{T}^{\text{g}}$
\begin{subequations}
\begin{align}
   & a^{\text{LCDF}}_{k\tau}+g_{k\tau}\leq U^{\text{inj}}_k & \label{ng-c3}
\end{align}
\end{subequations}

\bt{The NG and LCDF import limits are imposed in constraints~\eqref{ng-c3}}. 

\noindent\textbf{Flow Constraints:}
For every $\ell \in \mathcal{L}^{\text{g}}, \tau \in \mathcal{T}^{\text{g}}, j\in \mathcal{N}^s$
\begin{subequations}
\begin{align}
       & f^{\text{g}}_{\ell \tau } \leq I^{\text{pipe}}_{\ell}& \text{if } \mathcal{I}^{\text{pipe}}_{\ell}=1 \label{ng-c5}\\
    & f^{\text{g}}_{\ell \tau } \leq U^{\text{pipe}}_{\ell} z^{\text{g}}_{\ell}& \text{if } \mathcal{I}^{\text{pipe}}_{\ell}=0  \label{ng-c6}\\
    &\sum_{k \in \mathcal{N}^{\text{g}}:j\in \mathcal{A}^s_k} f^{\text{gl}}_{kj \tau} =s^{\text{liq}}_{j\tau} & \label{ng-c7}\\
    &\sum_{k \in \mathcal{N}^{\text{g}}:j\in \mathcal{A}^s_k } f^{\text{vg}}_{jk \tau}=s^{\text{vpr}}_{j\tau}  & \label{ng-c8}
\end{align}
\end{subequations}
The constraints~\eqref{ng-c5} and \eqref{ng-c6}  limit the flow between NG nodes for existing and candidate pipelines, respectively. The flow to liquefaction facilities is calculated in constraints~\eqref{ng-c7}. Similarly, the flow out of vaporization facilities is modeled via constraints~\eqref{ng-c8}. 

\noindent\textbf{Storage Constraints:} 
For every $j\in \mathcal{N}^s, \tau \in \mathcal{T}^{\text{g}}$
\begin{subequations}
\begin{align}
   &s^{\text{gStr}}_{j\tau} = (1-\beta) s^{\text{gStr}}_{j,\tau-1}+\gamma^{\text{liqCh}}_j s^{\text{liq}}_{j\tau}-\frac{s^{\text{vpr}}_{j\tau}}{\gamma^{\text{vprDis}}_j}  &  \label{ng-c9}\\
    &s^{\text{vpr}}_{j\tau}\leq  I^{\text{vpr}}_{j}+x^{\text{vpr}}_j  &  \label{ng-c10}\\
    &s^{\text{gStr}}_{j\tau}\leq  I^{\text{gStr}}_{j}+x^{\text{gStr}}_j  &  \label{ng-c11}
\end{align}
\end{subequations}
Constraints~\eqref{ng-c9} ensure the storage balance. Constraints~\eqref{ng-c10} and \eqref{ng-c11} limit the capacity of vaporization and storage tanks to their initial capacity plus the increased capacity, respectively.

\subsection{Coupling Constraints}
The following constraints are coupling constraints that relate decisions of the two systems.
\begin{subequations}
\begin{align}
    & \sum_{k \in \mathcal{A}^{\text{e}}_n} f^{\text{ge}}_{k n\tau } =   \sum_{t \in \mathfrak{T}^{\text{e}}_\tau} \sum_{i \in \mathcal{G}} h_i p_{nti} &\hspace{-2cm} n\in \mathcal{N}^{\text{e}}, \tau \in \mathfrak{R}\label{coup-1}\\
 &\mathcal{E}^{\text{e}} = \sum_{n\in \mathcal{N}^{\text{e}}}\sum_{t\in \mathcal{T}^e}\sum_{i \in \mathcal{G}} w_t(1-\eta_i)\eta^{\text{g}} h_i p_{nti}&\notag \\
 &\mathcal{E}^{\text{g}} =\sum_{k \in \mathcal{N}^g}\sum_{\tau \in \mathcal{T}^g} \eta^{\text{g}}( D^{\text{g}}_{k\tau}-  a^{\text{LCDF}}_{k\tau}-a^{\text{g}}_{k\tau})\notag \\
 &  \mathcal{E}^{\text{e}}+\mathcal{E}^{\text{g}} \leq (1-\zeta) (U^{\text{e}}_{\text{emis}} +U^{\text{g}}_{\text{emis}}) &\label{coup-2}
\end{align}
\end{subequations}
The first coupling constraints~\eqref{coup-1} captures the flow of NG to the power network for each node and at each time period. \bt{This constraint is the most common coupling constraint considered in the literature \citep{ChaudryEtal2014, ZhaoEtal2017, ShaoEtal2017,zhangEtal2018, VonWaldEtal2022}.}

The variable $\mathcal{E}^{\text{e}}$ accounts for CO$_2$ emission due to the consumption of NG in the power system. The variable  $\mathcal{E}^{\text{g}}$ computes the emission from NG system by subtracting the demand from LCDF consumption and gas load shedding. The second coupling constraint~\eqref{coup-2} ensures that the net CO$_2$ emissions associated with the power-gas system are below a pre-specified threshold value, which is defined based on a baseline emissions level. The first term is the emissions due to non-power NG consumption (i.e., NG consumption in the NG system such as space heating, industry use, and transportation). Since the model does not track whether LCDF is used to meet non-power  NG demand or for power generation, the first term computes gross emissions from all NG use presuming it is all fossil and then subtracts emissions benefits from using LCDF. 

Here we treat LCDF as a carbon-neutral fuel source \cite{ColeEtal2021}, and thus the combustion emissions associated with its end-use are equal to the emissions captured during its production.
{ }The second term captures the emission from NG-fired power plants.
Alternatively, the emission constraints can only be applied to the power system as in \cite{SepulvedaEtal2021} or separately applied to each system as in \cite{VonWaldEtal2022}. We evaluate the former approach by reformulating the emissions constraint to only consider power system emissions and explore the impact of such a constraint in Section~\ref{sec:results}. \bt{In practice, the operations of NG network, depend on the provision of electricity from power system (e.g. for powering gas compression stations). However, this is generally a relatively small fraction of total bulk electricity system demand, and thus we have ignored the electricity requirement to operate the NG system.}

\section{Case study and scenarios of interest} \label{sec:case-study}
\subsection{Case study}
We apply the proposed \ModelName model to study joint planning of power and NG systems for a case study based on the New England region in the year 2050. We approximate the region's power and NG networks using publicly available data (see Table~\ref{tab:input-data-summary}) while recognizing the need to choose the  model's spatial and temporal resolution to ensure reasonable computational run times.  The resulting power system consists of 6 nodes where each node represents a state of New England. There are 23 existing and \bt{7} candidate transmission lines between power network nodes. The power system has 5 existing and 7 new power plant types as described in Table~\ref{tab:exis_plant_params} and Table~\ref{tab:new_plant_params} respectively. The existing types include gas-fired (``ng''), solar, onshore wind (``wind''), hydropower (``hydro''), and nuclear. New plant types include NG combustion turbine (``OCGT''), combined cycle (``CCGT''), combined cycle with carbon capture and storage technology (``CCGT-CCS''), utility-scale solar (solar-UPV), onshore wind (``wind-new''), offshore wind (``wind-offshore''), and nuclear (``nuclear-new''). \bt{We assume that all thermal plants (i.e., ``ng'', ``OCGT'', ``CCGT'', ``CCGT-CCS'') can use LCDF and/or NG as their fuel.}

The existing power plant capacity is modeled based on resources deployed as of 2016 and we model the flexibility to retire these assets if deemed economic. We consider Li-ion battery (4 hours) as the available short-duration storage and metal-air battery as representative of emerging long-duration storage technologies (roughly classified as those that are capable of storing energy over periods greater than 12 hours).  
The NG network consists of 18 NG nodes and 7 SVL nodes. The NG nodes are connected via \bt{28 existing and 36 candidate pipelines}. The connections between NG nodes SVL nodes as well as NG nodes and power nodes are described in Fig.~\ref{fig:node-connections}. Table~\ref{tab:input-data-summary} summarizes the major data inputs into the model for characterizing the electric and NG system and associated technologies, with further details available in the \ref{App:input-data}. 
JPoNG codes as well as all data used in this study are available in a GitHub repository~\citep{GithubRepo}.

\begin{table*}[ht]
    \centering
    \caption{Summary of major data input, their values, and references}
    \label{tab:input-data-summary}
    \begin{tabular}{p{0.25\linewidth}  p{0.35\linewidth}  p{0.35\linewidth}}
    \toprule
\textbf{Data} & \textbf{Description} & \textbf{Notes} \\
\midrule
Annual discount rate &	7.1\% & 	applied to all investments in  the model uniformly \\
\midrule
\midrule
\multicolumn{3}{c}{\textbf{Power System}}\\

Power system resource capital costs &	Sourced for year 2050 – see Table~\ref{tab:exis_plant_params} &	Based on NREL Annual technology baseline \citep{ATB2021}\\
Electric transmission line investment costs &	3500 $\$$/MW/mile	&Based on Regional Energy Deployment System (ReEDS) \cite{ReEDS2019}\\
Electric network topology data &	Explained in \ref{app:power-data} &	Sourced mainly from \cite{PowerSimData}, \cite{ATB2021}, \cite{SepulvedaEtal2021}, \cite{ReEDS2019}\\
Electricity demand scenarios & 	Explained in \ref{app:electrification-scen} &	Based on NREL’s Electrification Future Study Load Profile dataset \cite{EFSload2022}\\
\midrule
\midrule
\multicolumn{3}{c}{\textbf{NG System}}\\
Gas system demand scenarios &	Explained in \ref{app:electrification-scen} &	Based on \cite{EFSload2022} and \cite{eiaWebsite2021}\\
Gas network topology data & 	Explained in \ref{app-NG-topology} &	Based on \citep{eiaWebsite2021}\\
Gas SVL nodes &	Explained in \ref{app-NG-topology} &	Sourced mainly from \cite{NGstatGuide2021}\\
Gas pipeline investment costs &5.34 million $\$$/mile	 &  Based on \citep{PipePrice2021} \\
NG import price &	5.45 $\$$/MMBtu & Represents cost of fossil NG injections into different nodes in the network. Based on \cite{SepulvedaEtal2021}\\
LCDF import price&	20 $\$$/MMBtu &	Represents cost of LCDF injection into the network. Based on \cite{ColeEtal2021}\\
CO$_2$ emissions constraints &	Explained in \ref{app:emission-amount} & 	Based on \cite{eiaCO2emis}\\
Case C1: power system only emissions limits&
35, 37, 39, 41 million tons for 80\% to 95\% decarbonization goals, respectively&\\
Case C2 and C3 (combined NG and power system emissions limits) & 
54, 57, 61, 64 million tons  for 80\% to 95\% decarbonization goal, respectively&\\
Emission factor &0.053 ton/MMBtu for NG, $\quad$ 0 for LCDF& from \cite{eiaWebsite2021}\\  
\bottomrule
    \end{tabular}
\end{table*}

\subsection{Scenarios of interest}
\begin{figure}
    \centering
    \includegraphics[width=0.45\textwidth]{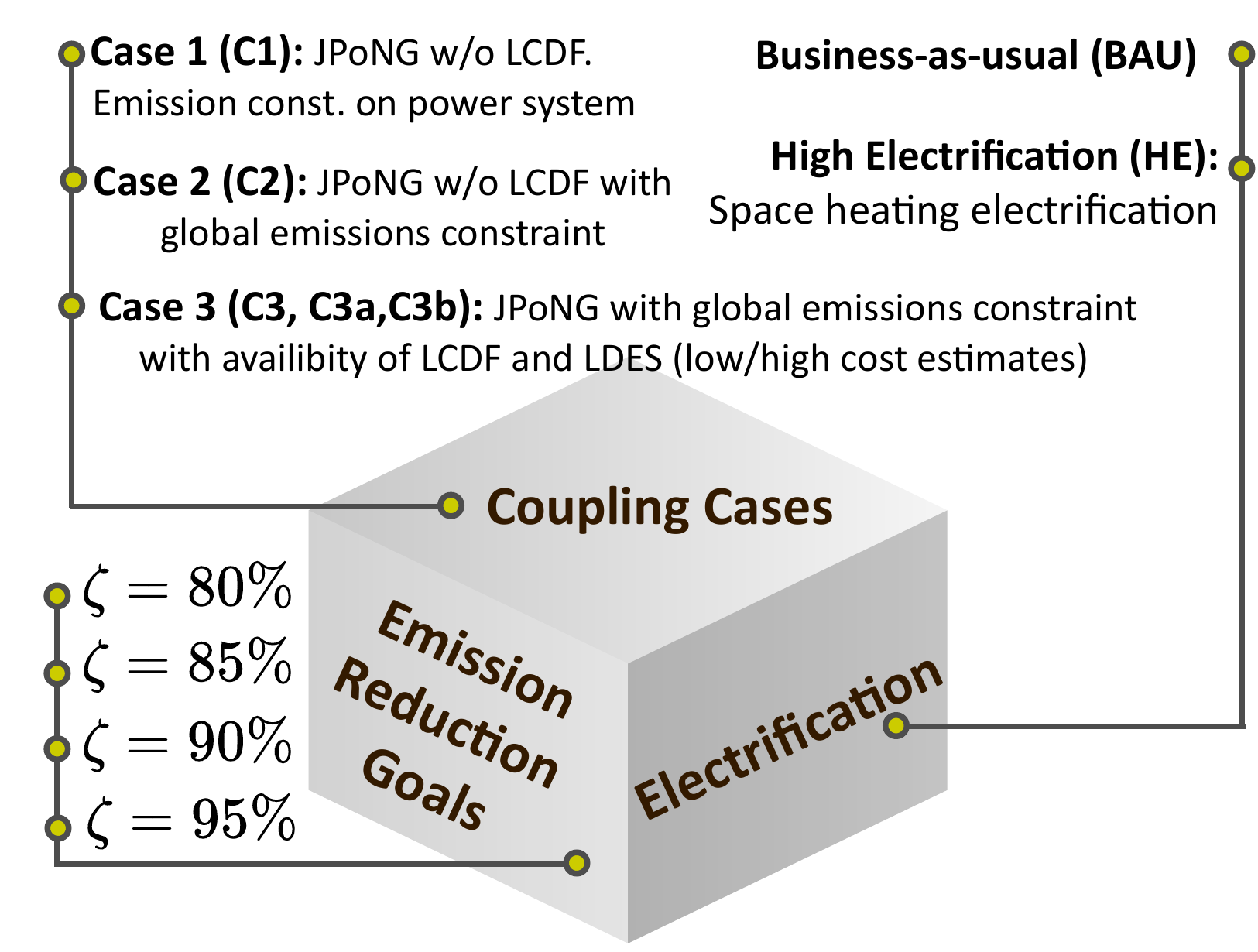}
    \caption{Technology cases, demand profiles, and CO$_2$ emissions reduction goals used in the computational experiments. LCDF = Low-carbon drop-in fuels. LDES = Long-duration energy storage. 
    }
    \label{fig:comp-design}
\end{figure}

\textbf{Coupling Cases:} we evaluate the model for three scenario sets or \textit{cases}, described in Fig.~\ref{fig:comp-design}, that capture the following technology and policy aspects: a) whether emissions constraint is imposed on power system or the combined system, b) end-use electricity and NG demand profiles and c) availability of low-carbon substitutes for NG, namely LCDF. We model the availability of LDES in cases C3a and C3b. For each case, we evaluate the model under four CO$_2$ emissions reduction goals, which are defined based on relative sectoral reduction compared to 1990 levels for the case study of interest \cite{eiaCO2emis}. \ref{app:emission-amount} provides more details.



Per scenario dimensions shown in Fig.~\ref{fig:comp-design}, case 1 (C1) considers \ModelName {} without LCDF or LDES and with emission constraint applied only on the power system. 
Hence, in C1, the Constraint~\eqref{coup-2} is replaced with:
$$
\mathcal{E}^{\text{e}} \leq (1-\zeta) U^{\text{e}}_{\text{emis}}
$$

Case 2 (C2) differs from C1 in that it considers a joint CO$_2$ emissions constraint for both systems as modeled by~\eqref{coup-2}, but still does not consider the availability of LCDF or LDES.
Finally, case 3 (C3) is the complete model described in Section~\ref{sec:model-formulation} in which a global emissions constraint and the possibility of LCDF import is considered. Additionally, we consider LDES in cases C3a and C3b with low and high estimates of investment and operational costs respectively.

 \textbf{Emissions reduction goals:} most New England states have goals for at least 80$\%$ economy-wide reduction in greenhouse gas (GHG) emission by 2050 vs. 1990 levels with some states targeting higher reductions \cite{MITEI2022FES}. Therefore, we consider four CO$_2$ emissions reduction scenarios  relative to the 1990 baseline, with reduction levels $\zeta$ which include 80$\%$, 85$\%$, 90$\%$, and 95$\%$.

\textbf{Demand characterization:} electrification of end-use technologies is a key aspect of most regional economy-wide decarbonization plans and can greatly impact energy infrastructure investment and emission trajectories \cite{EFS2021}. Heating electrification, in particular, is expected to play a key role in achieving emissions reduction goals in many parts of the US including New England \cite{HeatingNew2022}. Accordingly, we consider alternative end-use electricity and NG demand scenarios to study the impact of space heating electrification in residential and commercial building stock on the planning of power-NG infrastructure. We define business-as-usual (BAU) and high-electrification (HE) demand scenarios that assume electrification of space heating by replacing currently NG-fueled heaters with air- and ground-source heat pumps, albeit at different rates. The BAU scenario considers that minimum electrification is realized based on the current trend for the adoption of heat pumps. The HE scenario, however, accounts for the accelerated displacement of NG consumption and increasing power demand as a result of the mass adoption of heat pumps. 

The power system demand in both HE and BAU scenarios is based on the 2050 hourly load profiles provided as part of the NREL's Electrification Future Study (EFS) Load Profile dataset \cite{EFSload2022}. The repository contains hourly load profiles for various electrification (Reference, Medium, High) and technology advancement (Slow, Moderate, Rapid) scenarios. The load profiles are provided for several years including 2050 and are further disaggregated by state, sector (residential, commercial, etc.), and subsectors (space heating and cooling, water heating, etc.). For BAU scenario, we consider the aggregated state-level hourly demand profile for the \textit{Reference} electrification level with \textit{Moderate} technology advancement.

To isolate the effect of heating electrification, we consider the HE demand scenario that differs from the BAU scenario in the enhanced electrification of the space heating subsector. Specifically, 
we replace the  ``space heating and cooling'' subsector of the ``Reference'' and ``Moderate'' load profiles with ``High'' electrification and ``Moderate'' technology advancement load data, respectively. Once the space heating and cooling subsector are replaced, we aggregate based on sector and subsector to get state-level hourly electricity demand profile for the HE scenario. In the EFS scenarios, space heating comprises about 7$\%$ of the total power load in 2020, and its share increased slightly to 7.32\% under \textit{Reference} electrification level with \textit{Moderate} technology advancement, and to 13.21$\%$ under \textit{High} electrification level with \textit{Moderate} technology advancement. In our modeled electricity demand scenario, the total load of  BAU scenario is 128.64 TWh, of which 7.32$\%$ is incurred due to space heating. The load of the power system increase by 12.8$\%$ in HE scenario to 145.11 TWh with space heating making up 17.84 $\%$ of the total demand. 

We estimate the NG load displacement due to space heating electrification by the EFS data \cite{EFSload2022}, NREL's ``End-Use Load Profiles for the U.S. Building Stock'' project \cite{ResstockComstock2022}, and Energy Information Administration (EIA) \cite{eiaWebsite2021}. The details of the estimation is given in ~\ref{app:electrification-scen}. Overall, space heating electrification is estimated to reduce the total NG load by 13.9$\%$, from 5.15 $\times 10^7$ MMBtu in BAU to 4.44 $\times 10^7$ MMBtu in HE. Notice that the annual averaged efficiency of heat pumps is usually more than 100$\%$ \cite{MITEI2022FES}, so the absolute reduction of space heating in the NG load is greater than the increase in electricity demand.

\subsection{Numerical experiments}\label{subsec:num-expr}
All instances are implemented in Python programming language with Gurobi 9.5 solver and run the MIT Supercloud system with Intel Xeon Platinum 8260 processor with up to 48 cores
and 192 GB of RAM \cite{Supercloud2018}. We limit the CPU time to 10 hours  for all instances by which point all models were solved to optimality with mixed integer gap of 1\% or lower. We also set the value of $L^{\text{RPS}}$ to 50\% as suggested in \cite{E3Team2020}.

We select the representative days based on the approach proposed in \cite{BrennerEtal2022}. The method is developed for multivector energy systems such as the one we consider in this paper, and is based on physical topology and temporal demand pattern of the energy networks. We choose 30 representative days based on the results in ~\ref{app:rep-days} (see Fig.~\ref{fig:rep-costs}). The model complexity increase with the number of representative days, generation and storage technologies, and spatiotemporal resolution of the networks. 
For the given number of representative days, the resulting problem has more than 0.44 million constraints and 0.6 million variables of which about 400 are integer variables.

\section{Results}\label{sec:results}
\subsection {Joint planning with emissions limits on the power system }

\begin{figure*}[ht]
    \centering
    \includegraphics[width=0.95\textwidth]{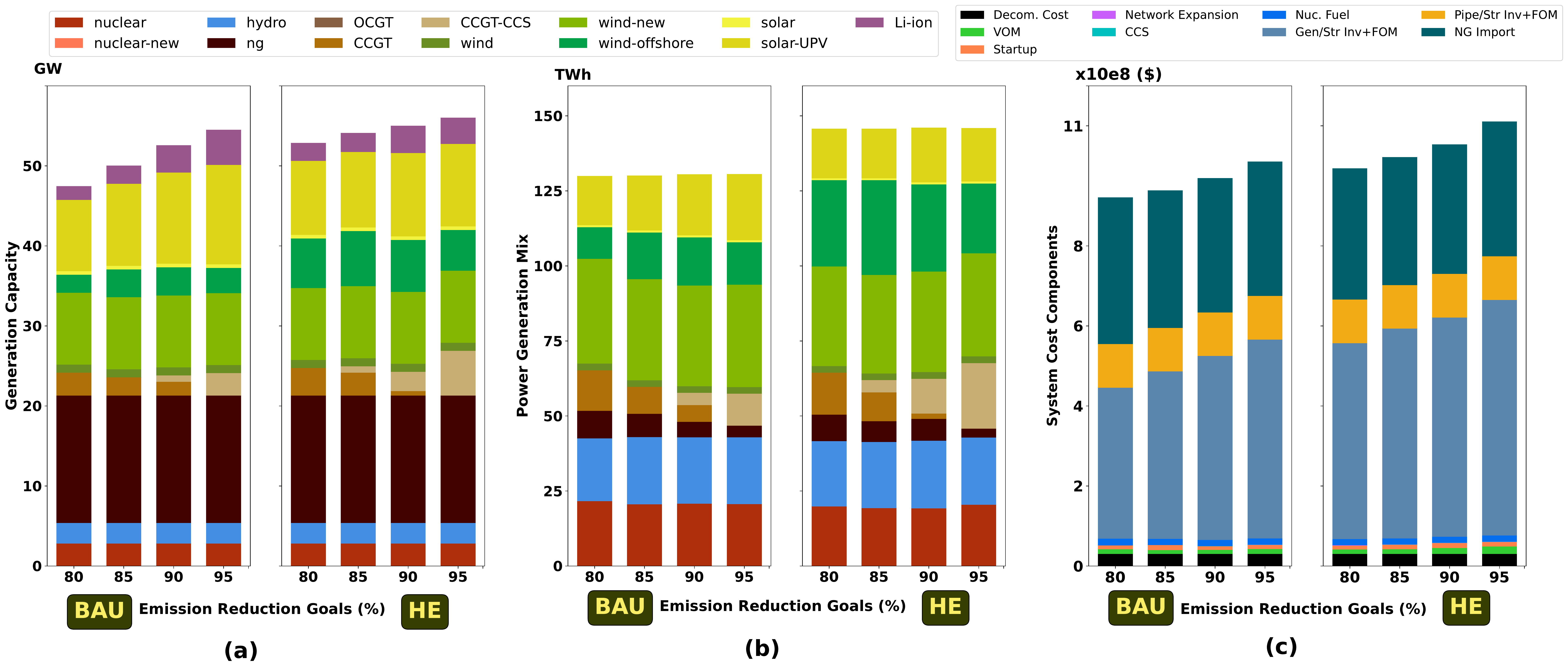}
    \caption{Capacity (a), generation (b) and system costs (c) for C1 under BAU and HE scenarios and different emissions reduction goals.\\
    The existing generation technologies include `ng'', ``wind'', ``hydro'' , ``solar'', and ``nuclear''.  The new generation technologies include ``OCGT'', ``CCGT'', ``CCGT-CCS'', ``wind-new'' (new onshore wind turbine), ``wind-offshore'', and ``solar-UPV'' (new solar panel). }
    \label{fig:case1-cap-gen-cost}
\end{figure*}

Fig.~\ref{fig:case1-cap-gen-cost} highlights the system outcomes and costs for the joint planning model instance with emissions constraints only on the power system (i.e., C1). Fig.~\ref{fig:case1-cap-gen-cost}a is consistent with other studies on power system decarbonization; we find that tightening the emissions constraint leads to increased deployment of low-carbon generation options, primarily VRE capacity supplemented with Li-ion storage capacity as well as CCGT-CCS (90\% CO$_2$ capture). Interestingly, all of the existing NG-fired power generation capacity is retained, albeit with much-reduced capacity utilization (3 to 7\% for BAU, and 2 to 6\% for HE scenario) as shown in Fig.~\ref{fig:case1-cap-gen-cost}b, \bt{which can be attributed to the high cost associated with decommissioning these plants.} 

This result further implies that it is more cost-effective to retain existing NG-fired plants rather than build and sparingly utilize additional new gas-fired plants without CCS which incurs additional investment costs. On the other hand, increasing electricity demand while meeting the same absolute emissions limit makes CCS-based power generation more cost-effective to deploy despite its higher heat rate and higher capital cost compared to CCGT without CCS. On average, this results in a \bt{32\%} increase in fuel (NG and LCDF) consumption in the power system between the HE and BAU scenarios. This increase is more than offset by the reduction in NG demand from buildings between BAU and HE scenarios for all emissions reduction goals where the total NG consumption in the HE scenario is \bt{ on average 5\% less} than the corresponding BAU scenario.

Across all the scenarios shown in Fig.~\ref{fig:case1-cap-gen-cost}, we see limited investment in network expansion costs, for both electricity and NG transport, corresponding to \bt{ 1 new transmission line and 2 pipelines} respectively. \bt{The capital investment cost in the power system (``Gen/Str inv+FOM'') makes up the bulk of the total system cost and is greater in the HE vs. BAU scenario, largely due to the need for additional low-carbon power generation capacity. The capital cost also increases as the emissions targets become more stringent in both scenarios owing to the need to overbuild VRE capacity relative to peak demand so as to provide sufficient VRE-based power during times of low VRE output (see Fig.~\ref{fig:case1-cap-gen-cost}a.) }

In addition, since much of the increased electricity demand originates from electrifying heating services in buildings during winter months, the HE scenario also results in greater deployment of wind, both onshore and offshore, as compared to solar PV (Fig.~\ref{fig:case1-cap-gen-cost}b). This is explained by the much lower solar resource availability during winter time, with average hourly capacity factors for the modeled representative days during winter months (Dec-Feb) below 13\%. 
Li-ion storage deployments are also reduced in the HE scenario vs. BAU scenario since it is generally cost-effective for short-duration energy storage applications (installed durations of storage are between 3 to 4.5 hours for scenarios shown in Fig.~\ref{fig:case1-cap-gen-cost}a) and strongly correlated with solar deployment. The additional electrification of building heat between HE and BAU scenarios results in modest increases in system costs ranging between 8-10\% as shown in Fig.~\ref{fig:case1-cap-gen-cost}c.

\subsection{Impact of global emissions constraint (C2 vs. C1)} \label{sec:c2VSc1}


Here, we explore the impact of another layer of realism in energy infrastructure planning, namely limiting emissions from both NG and electricity systems in a combined manner that is consistent with mid-century economy-wide decarbonization goals.

\begin{figure*}[ht]
    \centering
    \includegraphics[width=0.95\textwidth]{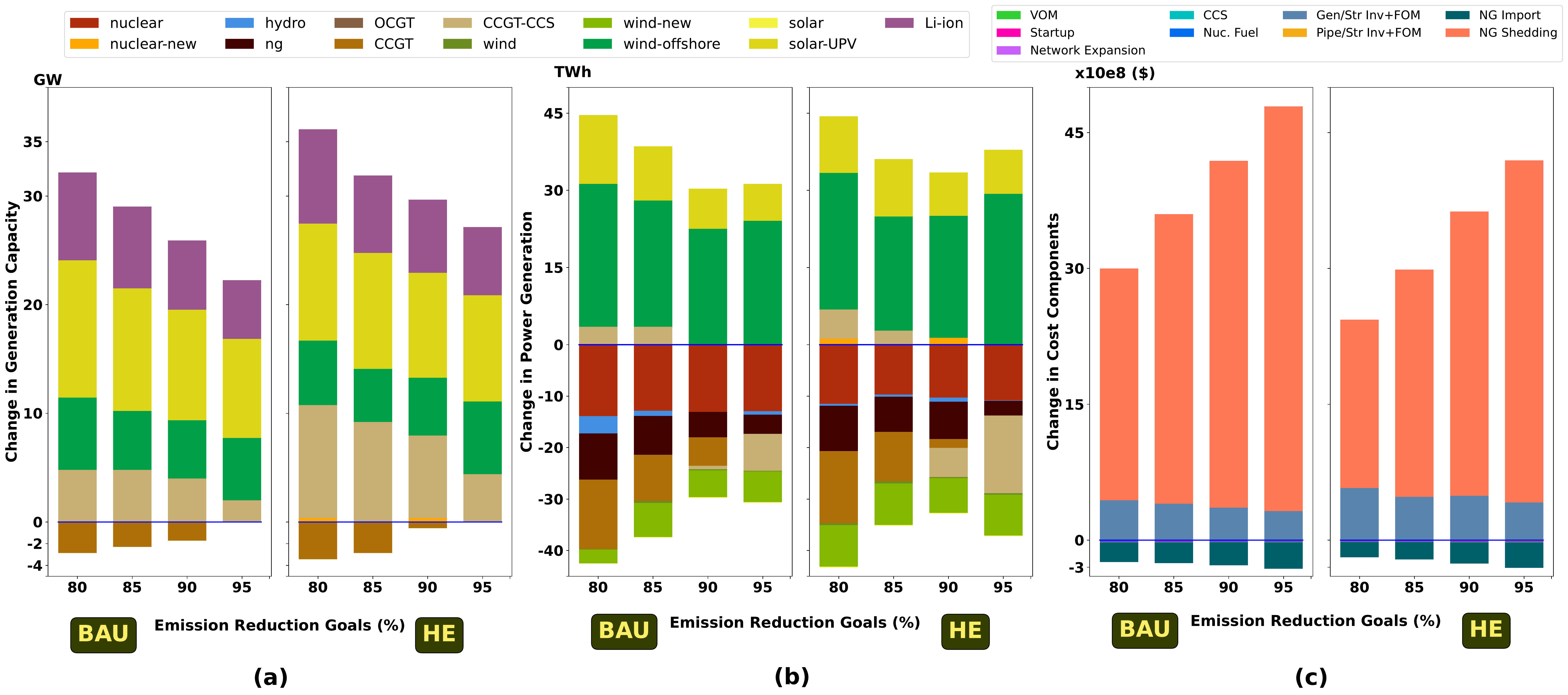}
    \caption{ Difference in capacity (a), generation (b), and system costs (c) between C1 and C2 ($C2-C1$)}
    \label{fig:diff-plot-cap-gen-cost-1vs2}
\end{figure*}

Fig.~\ref{fig:diff-plot-cap-gen-cost-1vs2} highlights the difference in power system capacity, electricity generation and total system cost between cases C2 and C1. Fig.~\ref{fig:diff-plot-cap-gen-cost-1vs2} shows that a global emissions constraint  disproportionately allocates emissions to meeting non-power NG demand while reducing NG-related emissions in the power system through increased VRE, short-duration storage and CCGT-CCS capacity deployment. For example, in the HE scenario, VRE capacity and CCGT-CCS capacity on average increase in  C2 relative to  C1 by \bt{58-68\%} and \bt{79-1150\%} respectively, while total fuel consumption in the power system drops by \bt{67-76\%}. Consequently, power system emissions in C2 are \bt{86-98\%} lower than sectoral emissions in  C1 across the BAU and HE scenarios, which highlights the importance for emissions trading across systems to meet the global emissions reduction goals. \bt{Interestingly, solar and onshore wind deployment limits for BAU and onshore wind limit for HE, obtained from other studies on regional decarbonization pathways \citep{E3Team2020}, are binding for all emissions goals across C2. This implies that further VRE expansion and reductions in NG use in the power system may be cost-effective, if deployment limits are relaxed.}

Increase in VRE capacity leads to  greater deployment of battery energy storage (both power and energy capacity) across all scenarios in Fig.~\ref{fig:diff-plot-cap-gen-cost-1vs2} that contribute to reducing VRE curtailment. The increased reliance on VRE also increases rated duration\footnote{rated duration is defined as energy capacity times discharge efficiency divided by power capacity} of energy storage capacity deployed from 2.96-3.82 hours in C1 to 3.92-4.23 hours in C2. 

Under the HE scenario, where non-power NG demand is reduced and so are the associated emissions, the model favors the deployment of additional CCGT-CCS based power generation in favor of CCGT that can be used sparingly to meet the higher electricity demand while still adhering to the imposed global emissions limits (Fig.~\ref{fig:diff-plot-cap-gen-cost-1vs2}b). Still, in the absence of alternative strategies such as the availability of LCDF for dealing with emissions from non-power NG consumption, the model is unable to meet a fraction of non-power  NG demand (42-85\% and 36-87\% of non-power NG demand across the HE scenario and BAU scenario, respectively) while adhering to the imposed emissions limits. \bt{This is despite  the increased reliance on VRE  over NG power generation in scenario C2 vs. C1 as shown in (Fig.~\ref{fig:diff-plot-cap-gen-cost-1vs2}b.)}

As Fig.~\ref{fig:diff-plot-cap-gen-cost-1vs2}c shows, due to heavy penalty imposed on non-served NG demand in the model (at a cost of 1000\$/MMBtu), the associated load shedding costs dominate the cost differences between C2 and C1, where we also see that HE scenarios result in greater power system investment and reduced NG import cost as compared to the BAU scenario.  It is important to note that unsatisfied NG demand at this scale is not a realistic scenario but we have included it as part of a systematic investigation of the impact of various supply and demand drivers impact power-NG infrastructure evolution.

\subsection{Impact of low-carbon, drop-in fuels (C3 vs. C2) and its interplay with heating electrification} \label{sec:c3VSc2}
Here, we consider how the availability of LCDF can impact power-NG infrastructure investment outcomes under carbon emissions constraints. \bt{As discussed earlier,  we model LCDF as a carbon-neutral fuel at a price that is four times that of NG at 20\$/MMBtu. Later on, we test the sensitivity of our results to varying LCDF prices in Section~\ref{ssec:sens-fuel-price}.}

\begin{figure*}[ht]
    \centering
    \includegraphics[width=0.95\textwidth]{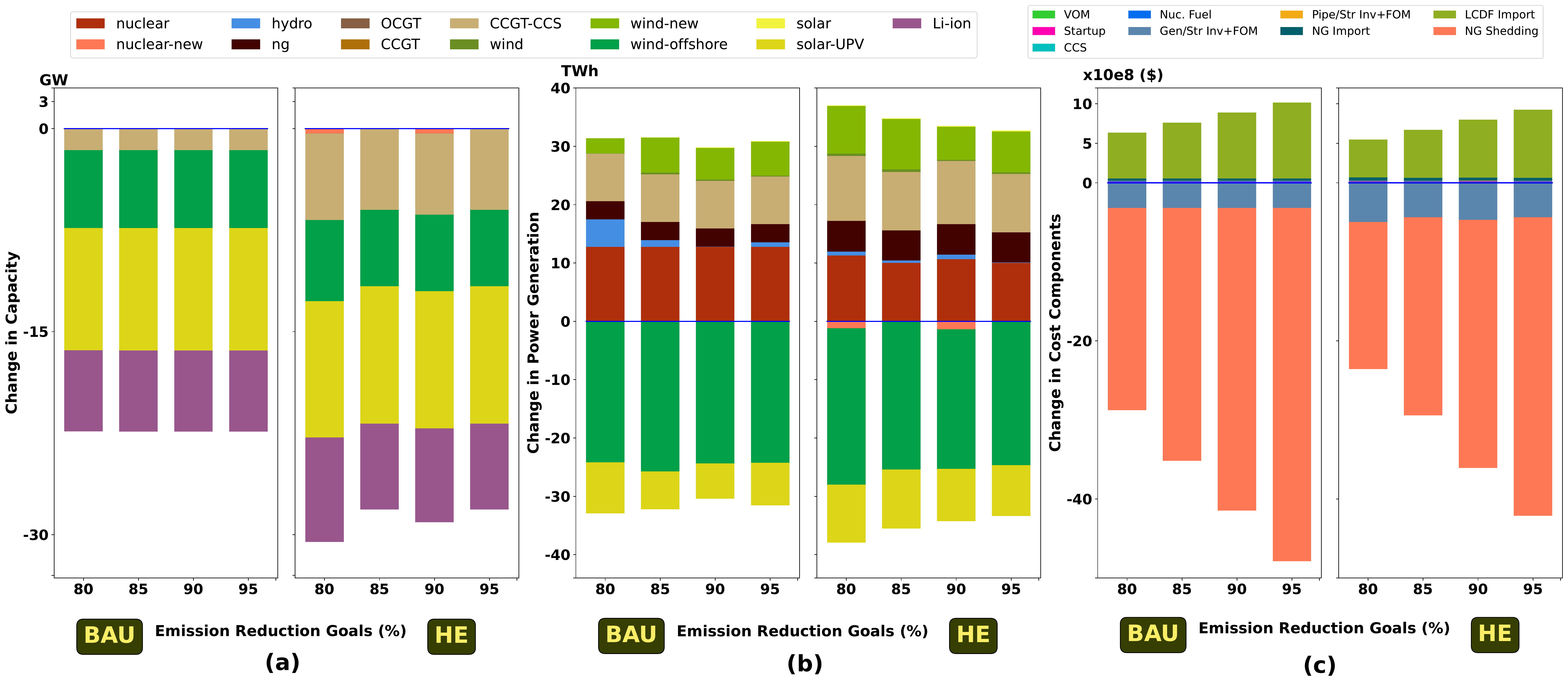}
    \caption{ Difference in capacity (a), generation (b) and system costs (c) between C2 and C3 scenarios ($C3-C2$) across various emissions reduction cases.}
    \label{fig:diff-plot-cap-gen-cost-2vs3}
\end{figure*}
\begin{figure*}[ht]
    \begin{minipage}{.5\textwidth}
        \includegraphics[width=\textwidth]{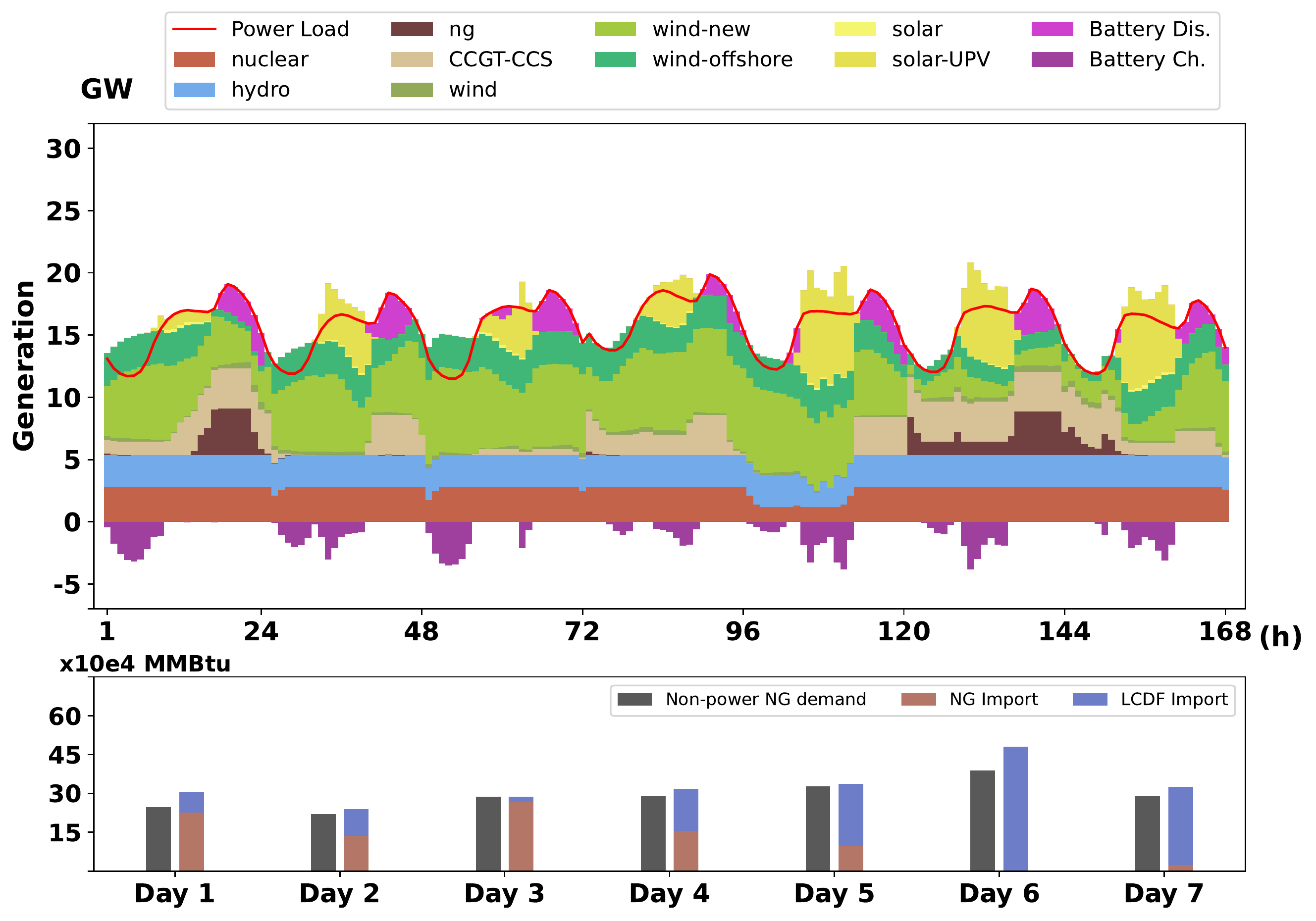}
    \captionof*{figure}{(a)}
        \label{fig:hourly-gen-BAU-winter-80}
    \end{minipage}%
    \begin{minipage}{.5\textwidth}
        \includegraphics[width=\textwidth]{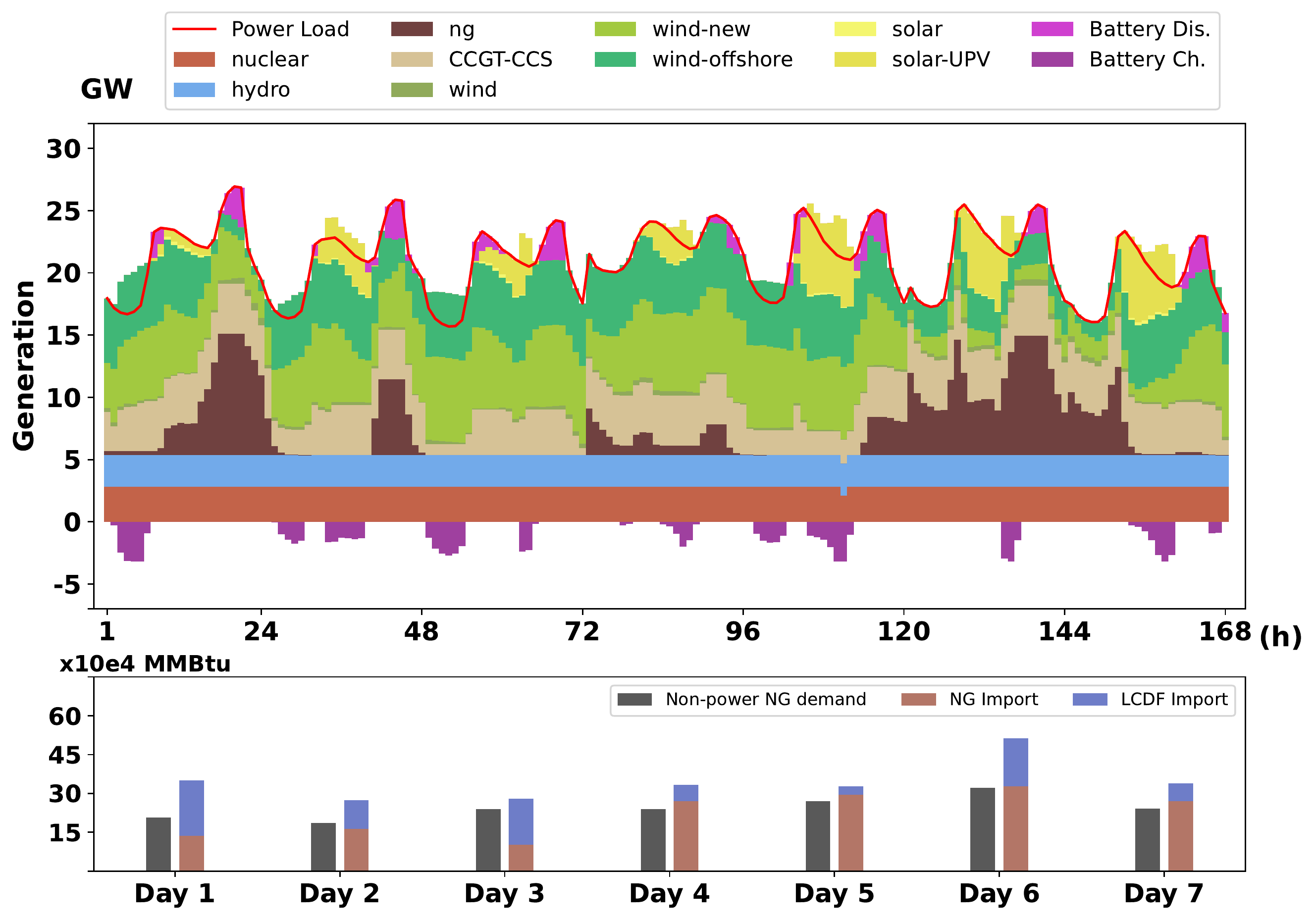}
        \captionof*{figure}{(b)}
        \label{fig:hourly-gen-HE-winter-80}
    \end{minipage}%
     \caption{Hourly mix for a typical winter week in C3, BAU (a) and HE (b) scenarios with $\zeta=80\%$. A Typical winter week is a set of seven representative days in winter months, so these days are not necessarily contiguous.}
        \label{fig:hourly-gen-winter-80}
\end{figure*}

Fig.~\ref{fig:diff-plot-cap-gen-cost-2vs3} shows that LCDF makes it possible to fully serve electricity and NG demand across all the emissions and demand scenarios, as seen by a reduction in NG load shedding related costs between C3 and C2. Compared to C2, C3 sees increased generation from existing NG capacity (without CCS) and nuclear capacity (Fig.~\ref{fig:diff-plot-cap-gen-cost-2vs3}b) where the former operates in a flexible manner to balance VRE integration and thus also enables increased utilization of the less flexible nuclear generation fleet (see Table~\ref{tab:new_plant_params} for flexibility parameters). This operational pattern is evident in Fig.~\ref{fig:hourly-gen-winter-80}a and \ref{fig:hourly-gen-winter-80}b that highlights the dispatch of the power system during the modeled representative days during winter months for BAU and HE scenarios, respectively. 

\bt{The ability to use existing NG plants in a low-carbon manner via using LCDF in scenario C3 reduces investments in additional low-carbon power generation technologies compared to C2 scenario (Fig.~\ref{fig:diff-plot-cap-gen-cost-2vs3}a). Consequently, C3 has reduced power system investment costs compared to C2 scenario, which are amplified in the HE scenario owing to increasing emissions budget for the power sector (see ``Gen/Str inv+FOM'' in Fig.~\ref{fig:diff-plot-cap-gen-cost-2vs3}c). } \bt{Similar to C2, the resource availability for onshore wind is exhausted. However, unlike C2, only 50\% of the maximum capacity limit for solar plants is utilized due to increased role of CCGT-CCS plants. The network expansion for both networks remains modest at 1 transmission line and 2 pipelines.}

Although the generation mix for both scenarios contains a significant amount of VREs (see \ref{app:gen-mix}, Fig.~\ref{fig:gen-mix1-2-3}), it is still necessary to dispatch from other plants.  \bt{The higher utilization rate for the existing plants, primarily  during times of peak demand (see Fig.~\ref{fig:hourly-gen-winter-80}), further reduced the establishment of Li-ion storage as well as CCGT-CCS capacity.} At the same time, the availability of a carbon-neutral fuel like LCDF in C3 enables greater utilization of CCGT-CCS resources as compared to the C2 scenario (\bt{52-59\% vs. 6-8\%}).

\begin{figure*}[ht]
    \centering
    \includegraphics[width=0.7\textwidth]{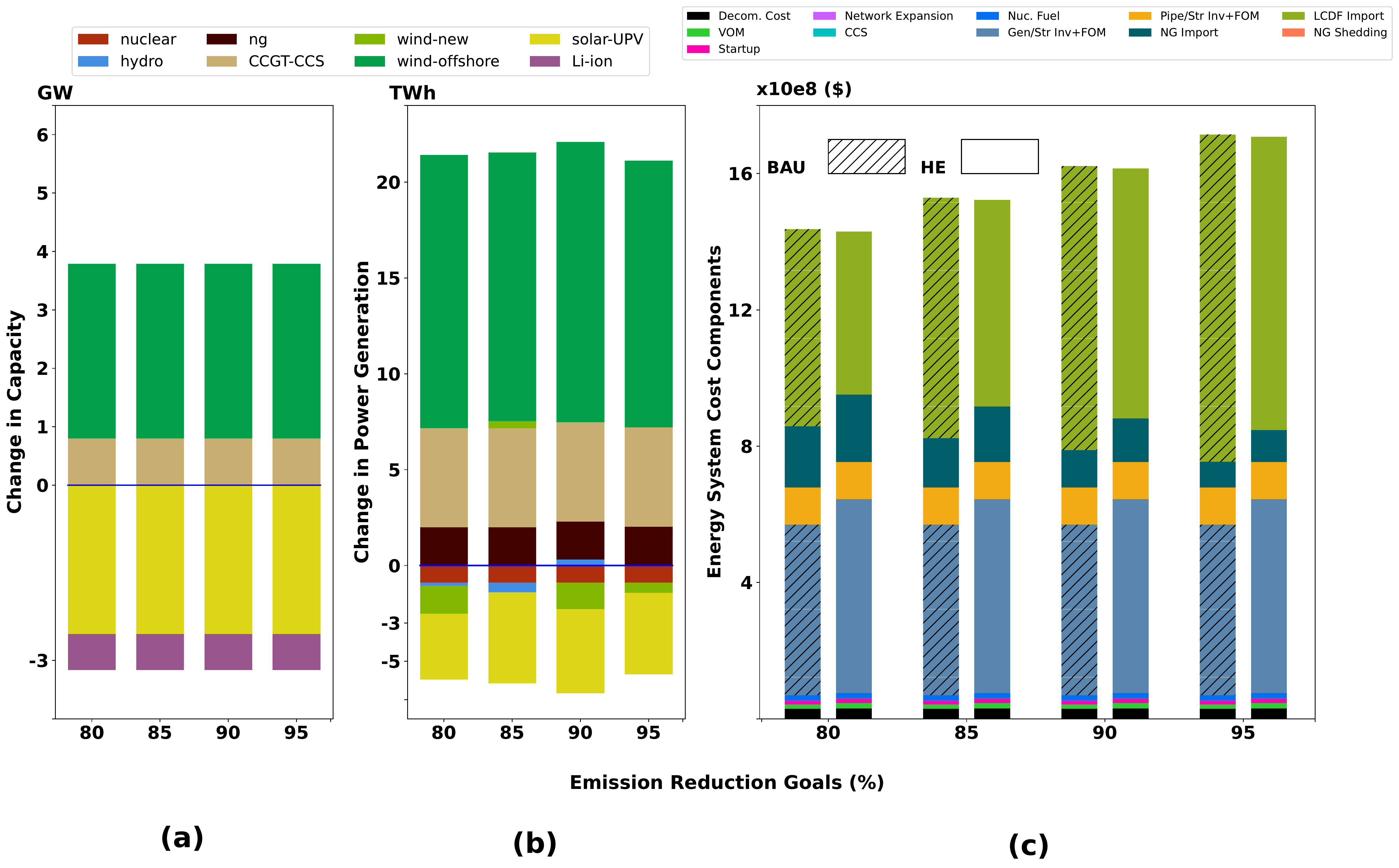}
    \caption{Difference in capacity (a) and generation (b) between BAU and HE (HE $-$ BAU) of C3. 
    Fig (c) shows system cost components for both BAU and HE scenarios of C3}
    \label{fig:case4-BAU-HE-cap-gen-cost}
\end{figure*}

Electrification of the buildings sector increases electricity demand in the winter months that results in a shift in the power system capacity mix to favor resources capable of producing electricity during those periods, namely wind generation and CCGT-CCS instead of solar PV (Fig.~\ref{fig:case4-BAU-HE-cap-gen-cost}). 
As noted in section~\ref{sec:c2VSc1}, Li-ion storage deployment complements solar PV deployment as compared to wind deployment and this explains why we see lesser Li-ion storage in the HE scenario. As highlighted in Fig.~\ref{fig:case4-BAU-HE-cap-gen-cost}c, the reduction in non-power NG demand in the HE scenario as compared to BAU scenario facilitates the use of additional NG in conjunction with CCS, in balancing the power system and therefore reduces the need to rely on more expensive fuels like LCDF for emissions abatement. Fig.~\ref{fig:hourly-gen-winter-80}b illustrates this point by highlighting a greater role for NG power generation during periods of peak electricity demand in the HE vs. BAU scenario. Across the emissions constraint scenarios, increasing end-use electrification results in \bt{10-17\%} lower overall LCDF use, while increasing overall NG consumption (both power and non-power) by \bt{10-24\%} and increasing power system investment costs by \bt{20\%}. 
\bt{ Note that the increase in NG consumption is feasible with the imposed emissions limits since NG is used primarily via ``CCGT-CCS'' plants that capture 90\% CO$_2$ and contribute to incremental electricity supply in the HE scenario.} Overall, the system cost for the HE scenario in C3 is on average \bt{0.4\%} lower compared to the BAU scenario suggesting the importance of electrification as a cost-effective mechanism for deep decarbonization of energy systems.

\subsection{Impact of Long-duration Energy Storage (C3a-b vs. C3)}

LDES is another emerging technology that is expected to play a major role in the energy transition \cite{SepulvedaEtal2021} and is typically characterized by much lower energy capital costs and round-trip efficiencies and higher power capital costs compared to Li-ion storage systems \cite{MITEI2022FES}. 

\begin{figure*}
    \centering
    \includegraphics[width=0.9\textwidth]{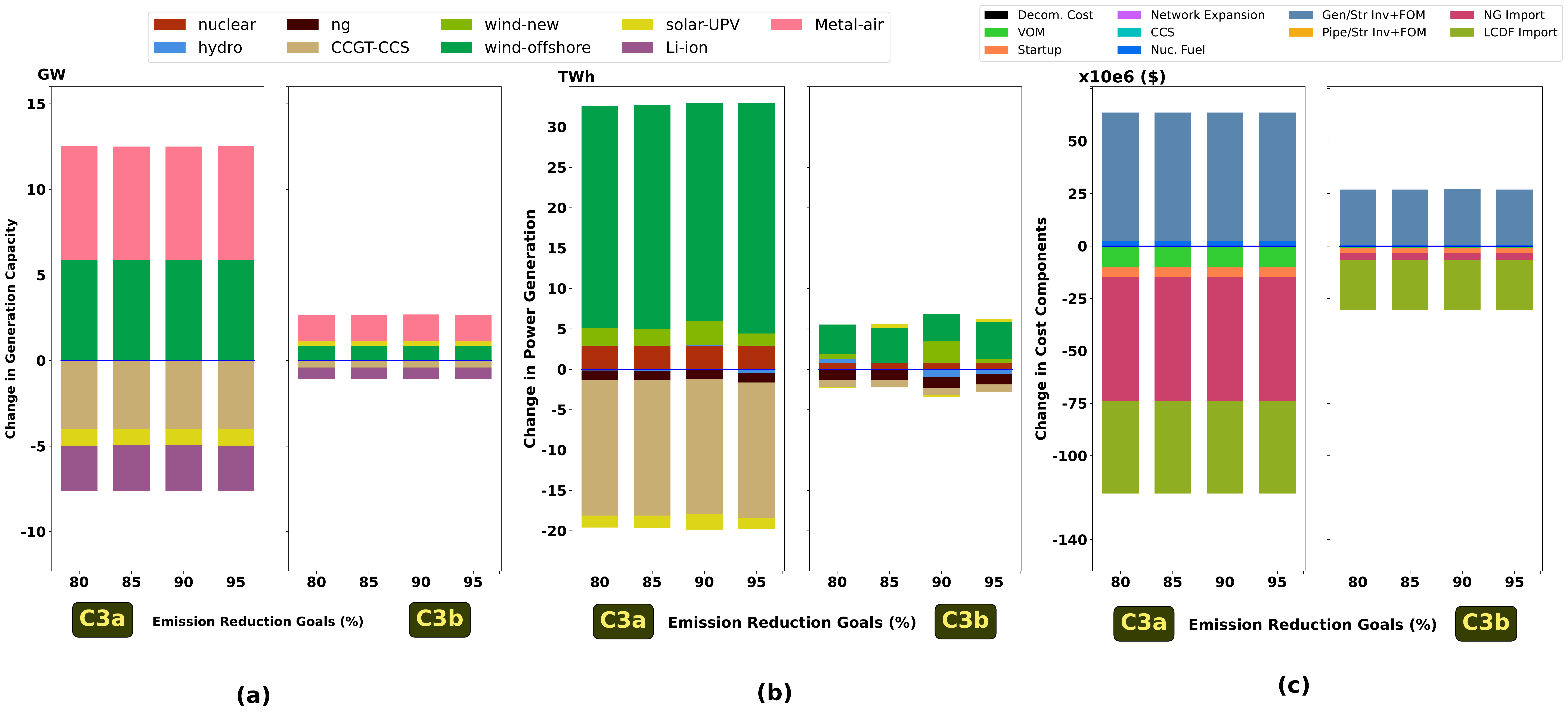}
    \caption{Difference in capacity (a), generation (b) and costs (c) between HE scenarios of C3 for metal-air with \textit{low, C3a} (C3a-C3) and \textit{high, C3b}  (C3b-C3) cost estimates}
    \label{fig:metal-air-cap-gen-cost}
\end{figure*}

While there are many alternative configurations of LDES systems being developed based on thermal, electrochemical, chemical and mechanical energy storage concepts, here we evaluate the impact of LDES technology participation in deeply decarbonized energy systems based on plausible low-cost (C3a) and high-cost (C3b) estimates for aqueous metal-air battery sourced from the literature \cite{MITEI2022FES}. For the HE demand scenario, Fig.~\ref{fig:metal-air-cap-gen-cost} compares outcomes between C3 with C3a and C3b, where the latter two correspond to scenarios with low (3a) and high (C3b) cost estimates for metal-air battery.

Fig.~\ref{fig:metal-air-cap-gen-cost}a and Fig.~\ref{fig:metal-air-cap-gen-cost}b show that the availability of LDES generally shifts the power system portfolio to rely on wind over CCGT-CCS, solar-UPV and short-duration storage, with the most pronounced impacts seen in the low-cost LDES scenario. \bt{Quantitatively, the fuel (NG and LCDF) consumption  is 11-31\% and 78-95\% less in C3a and C3b compared to C3.}  
The reduced use of fuel in the power system resulting from LDES adoption reduces the extent of coupling between electricity and NG infrastructure and also reduces the reliance on expensive fuels like LCDF for meeting non-power NG demand as depicted in Fig.~\ref{fig:metal-air-cap-gen-cost}c. Collectively, these factors result in LDES contributing to system cost reductions ranging between \bt{3.2-4.6\%} in the low-cost LDES scenario and \bt{0.2-0.3\%} in the high-cost LDES scenario shown in Fig.~\ref{fig:metal-air-cap-gen-cost}.


\subsection{Fuel Price sensitivity analysis} \label{ssec:sens-fuel-price}
\begin{figure*}
    \centering
    \includegraphics[width=\textwidth]{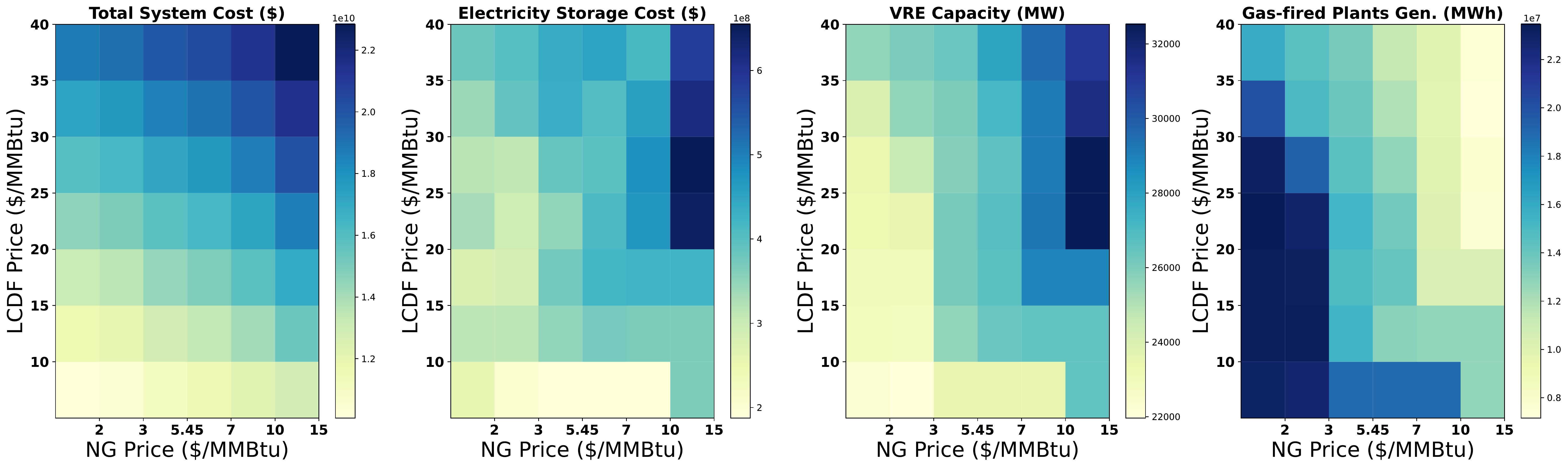}
    \caption{Sensitivity of planning outcomes to various LCDF and NG price combinations for BAU scenario under 80\% emissions reduction goal. The NG and LCDF prices are experimented at 6 and 7 levels, respectively.}
    \label{fig:fuel-price-sens}
\end{figure*}

\bt{Given the prominent role of fuel prices in the total system cost for the integrated energy system scenarios (C3), this section analyzes the sensitivity of the planning outcomes toward LCDF and NG prices. Fig.~\ref{fig:fuel-price-sens} highlights the impact of varying NG and LCDF prices on the total system cost, electricity storage costs, VRE capacity and gas-fired power generation under BAU scenario and 80\% emissions reduction goal. 
}

\bt{Fig.~\ref{fig:fuel-price-sens}  highlights the interaction effects between LCDF and NG price on system outcomes for total system cost and various system outcomes. Total system cost is more sensitive to LCDF cost; increasing NG price from 2 to 10 \$/MMBtu only increases total system cost by 22\%, 29\% and 27\% when LCDF price is fixed at 40, 20, and 10 \$/MMBtu. On the other hand, The increase in total system cost is 73\%, 79\% and 85\% when LCDF cost increases from lower to highest price and the NG price is fixed at 2, 5.45, and 15 \$/MMBtu, respectively. 
Compared to the base system cost at default fuel prices, the total system cost decreases by 30\% in the lowest levels of fuel prices, and increases by 59\% in the highest fuel price levels.} 

\bt{The installed capacity of VRE plants is sensitive to price of both fuels, especially NG price. The storage cost  exhibits the same sensitivity as VRE capacity indicating its dependence on the intermittent resources in the generation portfolio. The storage cost and VRE capacity are at their highest when NG price is at its highest and LCDF price is 50\% costlier than the base case at 30\$/MMBtu. In contrast, the power generation from gas-fired plants decreases as the price of LCDF and NG increases, and it is relatively high when either of the fuels is at its lowest price level. Compared to the default values, the generation from gas-fired plants increases by 54\% in the lower fuel prices, and decreases by 50\%  in the highest fuel prices. }


\section{Discussion and Concluding Remarks}\label{sec:conclusion}
Deep decarbonization of energy systems requires infrastructure planning whose scope needs to span across sectors and infrastructure for different vectors while considering increasing spatial and temporal variability of energy supply and demand and cross-sectoral interactions. 
In this paper, we developed an integrated optimization model for joint planning of power and NG infrastructure that considers intra-annual operation of these systems as well as a range of low-carbon energy technologies such as VRE generation, short and long-duration storage, LCDF, power generation with CCS. Importantly, our model allows for different temporal resolutions of operation for electric and NG infrastructure and facilitates the assessment of various coupling mechanisms between the two infrastructures including, emissions reduction goals, building stock electrification, as well as the use of NG and LCDF across the two systems.  

This framework reveals a number of key observations pertaining to deep decarbonization of energy systems and the interplay of technology and policy levers. \textbf{First}, similar to other studies on power system decarbonization \cite{JayadevEtal2020}, we find that cost-effective pathways for deep decarbonization of energy systems require aggressive expansion of VRE and energy storage systems, with a specific emphasis on onshore and offshore wind generation for the New England region. 

\textbf{Second}, a global emissions reduction viewpoint, enabled by our multi-vector modeling formulation, points to the cost-effectiveness of a disproportionate reduction in emissions from the power system rather than other sectors using NG for final energy. This is primarily due to the relatively low abatement cost resulting from the projected costs of VRE electricity generation options. For example, in both reference and high electrification scenarios of C3, gross emissions from the power system represent \bt{14\% and 23\%} of the emission budget, respectively. 

\textbf{Third}, across the scenarios we see reliance on a portfolio of strategies for system balancing and emissions reduction including VRE complemented with short and long-duration storage, dispatchable generation using both fossil fuel and more expensive LCDF as well as CCS, VRE curtailment and non-power NG demand reduction via electrification. The results also highlight a limited expansion of NG infrastructure, with at most one to three pipelines added across the scenarios, suggesting that any significant investments in the NG infrastructure may not be economical under future carbon-constrained scenarios.

\textbf{Fourth}, our modeling approach also helps in evaluating the relative cost impact of demand and supply-side measures. For the modeled high electrification levels,  which reduce non-power NG demand and increase the share of VRE generation, system costs are generally lower or similar compared to the BAU demand scenario. Counterintuitively, under the global emissions constraints and with the availability of LCDF, the total NG consumption in the HE scenario is \bt{10-24\%} greater in C3 than the corresponding BAU scenario, where the increased NG consumption is primarily directed towards CCS-based power generation. Furthermore, our analysis shows that the availability of low-cost LDES can considerably reduce the total system cost. 

\textbf{Finally}, we observe that for either electrification scenario, the total cost in many instances is in the same range suggesting that other metrics considered here such as the share of VRE sources, as well as other metrics not considered here such as land impacts, reliability measures, and robustness to inter-annual variations, might be better criteria to distinguish the importance of each pathway.

In terms of next steps, the proposed modeling framework has certain limitations and can be refined in multiple ways. First, the supply patterns for gas and LCDF can be better described by considering their spatio-temporal availability and price volatility as a function of consumption (e.g. similar to price volatility seen for NG \cite{GasPriceVolit2022}). Second, the extent of electrification of non-power NG demand is relatively low compared to state-level policy ambitions and future work could focus on developing granular demand scenarios for building, transportation and industry that are aligned with stated policy goals and evaluating their supply-side impacts. Third, we do not model the strategic retirement of NG pipeline \& storage infrastructure assets which could form an important component of NG system's cost if the displacement of gas with electricity is significant. \bt{Fourth, our model does not include any demand response (DR) programs, nor does it disaggregate the power and NG demand to various sectors.  DR programs have the potential to adjust the pattern of power demand, instead of supply, by either shifting energy  away from peak hours to certain periods or curtailing bulk power system demand through the adoption of distributed energy resources or voluntary reduction \citep{GhavidelEtal2020, MoralesEtal2022, SchittekatteEtal2022}. Accordingly, DR can play a major role in future decarbonized energy systems. It is worth noting that the role of DR can particularly be prominent when coupled with multi-energy systems such as district heating \citep{GuelpaVerda2021}.}

Apart from application details, future work can identify sources of uncertainty and include stochasticity in the modeling framework. Furthermore, the resiliency of the planning outcomes toward inter-annual demand and supply variations as well as extreme weather conditions is another direction to pursue. Finally, future studies can leverage decentralized planning approaches to better reflect the energy planning process that is usually carried out by multiple stakeholders in each energy system who usually have competing interests and sometimes opposing objectives.

\section{Nomenclature} \label{app:nomenclature}

\tablefirsthead{&\multicolumn{1}{c}{} \\ }
\tablehead{%
\\&\multicolumn{1}{c}{}\\ }
\tabletail{%
}
\tablelasttail{%
}
\small 
\begin{supertabular}{l l}
    \toprule
      \multicolumn{2}{l}{{\textbf{Sets}}}\\
      \midrule
    $\mathcal{N}^{\text{e}}$ &\hspace{-0.4cm} Power system nodes\\
    $\mathcal{P}$ &\hspace{-0.4cm}  Power plant types 
    \\
    $\mathcal{R} \subset \mathcal{P}$ &\hspace{-0.4cm} VRE power plant types 
    \\
    $\mathcal{G} \subset \mathcal{P}$ &\hspace{-0.4cm} gas-fired plant types 
    \\
    $\mathcal{CCS} \subset \mathcal{P}$ &\hspace{-0.4cm} gas-fired plant types with carbon capture technology
    \\
    $\mathcal{H} \subset \mathcal{P}$ &\hspace{-0.4cm} Thermal plant types  
    \\
    $\mathcal{Q} \subset \mathcal{P}$ &\hspace{-0.4cm} Technology with a resource availability limit\\
    $\mathcal{Q}'  $ &\hspace{-0.4cm} Set of technologies with \\
    &\hspace{-0.2cm} resource availability limits \\
    $\mathcal{T}^{\text{e}}$&\hspace{-0.4cm} Index set of representative hours for  power system\\
    $\mathfrak{R}$&\hspace{-0.4cm} Representative days\\
    $\mathfrak{T}^{\text{e}}_\tau$&\hspace{-0.4cm} Hours in $\mathcal{T}^{\text{e}}$ that are represented by day $\tau$\\
    $t^{\text{start}}_\tau,t^{\text{end}}_\tau$&\hspace{-0.4cm} First and last hour in $\mathfrak{T}^{\text{e}}_\tau$\\
     $\mathcal{L}^{\text{e}}$ &\hspace{-0.4cm} Existing and candidate transmission lines\\
     $\mathcal{L}^{\text{e}}_{nm}$ &\hspace{-0.4cm} Existing and candidate transmission lines\\
    $\mathcal{S}^{\text{eS}}_n$ &\hspace{-0.4cm}  Short-duration energy storage system types \\
    $\mathcal{S}^{\text{eL}}_n$ &\hspace{-0.4cm}  Long-duration energy storage system types\\
    $\mathcal{S}^{\text{e}}_n$ &\hspace{-0.4cm}  All energy storage systems types\\
    $\mathcal{A}^{\text{g}}_n$ &\hspace{-0.4cm} Adjacent NG  nodes for node  $n$\\
    &\hspace{-0.2cm} between node $n$ and $m$\vspace{0.1cm}\\ 
    \hdashline \vspace{-0.3cm}\\
    $ \mathcal{N}^{\text{g}}, \mathcal{N}^s$ &\hspace{-0.4cm} NG and SVL nodes\\
     $\mathcal{T}^{\text{g}}$&\hspace{-0.4cm} Days of the planning year\\
    $\mathcal{A}^s_k$ &\hspace{-0.4cm} Adjacent SVL facilities of node  $k$\\
$\mathcal{L}^{\text{g}}$ &\hspace{-0.4cm} Existing and candidate pipelines\\
    $\mathcal{L}^{\text{gExp}}_{k}$ &\hspace{-0.4cm} Existing and candidate pipelines\\
    & starting from node $k$\\
    $\mathcal{L}^{\text{gImp}}_{k}$ &\hspace{-0.4cm} Existing and candidate pipelines ending at node $k$\\
&\\
\midrule
    \multicolumn{2}{l}{{\textbf{Indices}}}   \\
    \midrule 
    $n,m$ & Power system node \\
    $k$ & NG system node\\
    $j$ & SVL facility node\\ 
    $i$ & Power generation plant type\\
    $r$ & Storage type for power network\\
    $\ell$& Electricity transmission line or gas pipeline\\
    $t$ & Time step for power system's operational periods\\
    $\tau$ & Time step for NG system's operational periods \\
&\\
\midrule
          \multicolumn{2}{l}{{\textbf{Annualized Cost Parameters}}}\\
      \midrule
         $C^{\text{inv}}_{i}$ &\hspace{-0.5cm}  CAPEX of plants, [$\$$/plant] \\
        $C^{\text{dec}}_i$ &\hspace{-0.5cm}  Plant decommissioning cost, [$\$$/plant]\\
         ${C}^{\text{trans}}_{\ell}$ &\hspace{-0.5cm} Transmission line establishment cost, [$\$$/line] \\
        $C^{\text{EnInv}}_{r}$ &\hspace{-0.5cm} Storage establishment energy-related cost, [$\$$/MWh]\\
      $C^{\text{pInv}}_{r}$ &\hspace{-0.5cm}  Storage establishment power-related cost, [$\$$/MW]\vspace{0.1cm}\\
     \hdashline \vspace{-0.3cm}\\
     
      ${C}^{\text{pipe}}_{\ell}$ &\hspace{-0.5cm}  Pipelines establishment cost, [$\$$/line] \\
      $C^{\text{strInv}}_{j}$ &\hspace{-0.5cm}  CAPEX of storage tanks at SVLs, [$\$$/MMBtu]\\
      $C^{\text{vprInv}}_{j}$ &\hspace{-0.5cm}  CAPEX of vapor. plants at SVLs, [$\$$/MMBtu/hour]\\
&\\
\midrule
          \multicolumn{2}{l}{{\textbf{Annual Costs}}}\\
      \midrule
         $C^{\text{fix}}_{i}$ &\hspace{-0.5cm} Fixed operating and maintenance \\
         & cost (FOM) for plants, [$\$$] \\
        $C^{\text{EnFix}}_{r}$ &\hspace{-0.5cm} Energy-related FOM for storage, [$\$$/MWh]\\
        $C^{\text{pFix}}_{r}$ &\hspace{-0.5cm} Power-related FOM for storage, [$\$$/MW] \vspace{0.1cm}\\  \hdashline \vspace{-0.3cm}\\  
      $C^{\text{strFix}}_{j}$ &\hspace{-0.5cm} FOM for storage tanks, [$\$$/MMBtu]\\
      $C^{\text{vprFix}}_{j}$ &\hspace{-0.5cm} FOM for vaporization plants, [$\$$/MMBtu/hour]\\
&\\
\midrule
          \multicolumn{2}{l}{{\textbf{Other Cost Parameters}}}\\
      \midrule
        $C^{\text{var}}_{i}$ &\hspace{-0.5cm} Variable operating and maintenance\\
        & cost (VOM) for  plants, [$\$$/MWh]\\
    $C^{\text{startUp}}_i$ &\hspace{-0.5cm} Start-up cost for plants, [$\$$]\\
     $C^{\text{eShed}}$ &\hspace{-0.5cm} Unsatisfied power demand cost, [$\$$/MWh] \\
      $C^{\text{fuel}}_i$ &\hspace{-0.5cm} Fuel price for plants, [$\$$/MMBtu]\vspace{0.1cm}\\  \hdashline \vspace{-0.3cm}\\
     $C^{\text{ng}}$ &\hspace{-0.5cm} Fuel price for NG, [$\$$/MMBtu]\\
     $C^{\text{LCDF}}$ &\hspace{-0.5cm} Price of LCDF, [$\$$/MMBtu]\\
     $C^{\text{gShed}}$ &\hspace{-0.5cm} Unsatisfied NG demand cost [$\$$/MMBtu]\\
&\\
\midrule 
          \multicolumn{2}{l}{{\textbf{Other Parameters for the Power System}}}\\
      \midrule
         $\rho_{nti}$ &\hspace{-0.5cm} Capacity (availability) factor for renewable plants\\
         $D^{\text{e}}_{nt}$ &\hspace{-0.5cm} Power demand, [MWh]\\
       $h_i$ &\hspace{-0.5cm} Heat rate, [MMBtu/MWh]\\
       $b_{\ell}$ &\hspace{-0.5cm} Susceptance of line $\ell\in \mathcal{L}^{\text{e}}$\\
        $\eta_i$ &\hspace{-0.5cm} Carbon capture rate, [\%]\\
       $U^{\text{prod}}_{i}$ &\hspace{-0.5cm} Nameplate capacity, [MW]\\
        $L^{\text{prod}}_{i}$ &\hspace{-0.5cm} Minimum stable output, [$\%$]\\
        $U^{\text{ramp}}_{i}$ &\hspace{-0.5cm} Ramping limit, [$\%$]\\
        $\gamma^{\text{eCh}}_r,\gamma^{\text{eDis}}_r$ &\hspace{-0.5cm} Charge/discharge rate for storage\\
        $\gamma^{\text{loss}}_r$ &\hspace{-0.5cm} hourly self-discharge rate for storage\\
        $I^{\text{trans}}_{\ell}$ &\hspace{-0.5cm} Initial capacity for transmission line $\ell$, [MW]\\
        $U^{\text{trans}}_\ell$ &\hspace{-0.5cm} Upper bound for capacity of transmission\\
        & line $\ell$, [MW]\\
        $\mathcal{I}^{\text{trans}}_{\ell}$ &\hspace{-0.5cm} 1, if  trans. line $\ell$ exists; 0, otherwise\\
        ${I}^{\text{num}}_{ni}$ &\hspace{-0.5cm} Initial number of plants\\
        $U^{\text{e}}_{\text{emis}}$ &\hspace{-0.5cm} Baseline emission of CO$_2$ in 1990 \\
        &from generation consumption, [ton]\\
       $U^{\text{CCS}}$ &\hspace{-0.5cm} Total annual carbon storage capacity, [ton]\\
        $L^{\text{RPS}}$ &\hspace{-0.5cm} Renewable Portfolio Standard (RPS) value\\
        $d_n$ &\hspace{-0.5cm} Distance between node $n$ and CO$_2$ storage site \\
        $E^{\text{pipe}}$ &\hspace{-0.5cm} Electric requirement for CO$_2$ pipeline\\
        & operations [MWh/mile/ton/hour]\\
        $E^{\text{pump}}$ &\hspace{-0.5cm} Electric requirement for compression of CO$_2$\\
        $E^{\text{cprs}}$ &\hspace{-0.5cm} Number of compressors required in \\
        & the pipeline from node $n$ to the storage site\\
        &  pipelines [MWh/ton/hour]\\
        $U^{\text{prod}}_{\mathcal{Q}}$ &\hspace{-0.5cm} Production capacity for set \\
        & of plants $\mathcal{Q}\subset \mathcal{P}$, [MW]\\
        $\zeta$ &\hspace{-0.5cm} emissions reduction goal\\
        $w_t$ &\hspace{-0.5cm} Weight of the representative period $t$\\
        $\phi^{\text{e}}_t$&\hspace{-0.5cm} Mapping of representative period $t$ to its\\
        & original period in the time series\\
&\\
\midrule 
          \multicolumn{2}{l}{{\textbf{Other Parameters for the NG System}}}\\
      \midrule
         $D^{\text{g}}_{k\tau}$ &\hspace{-0.5cm}  NG demand, [MMBtu]\\
        $\eta^{\text{g}}$ &\hspace{-0.5cm} Emission factor for NG [ton CO$_2$/MMBtu]\\
        $U^{\text{inj}}_k$ &\hspace{-0.5cm} Upper bound for NG supply, [MMBtu]\\
        $\gamma^{\text{liqCh}}_j$ &\hspace{-0.5cm} Charge efficiency of liquefaction plant\\
        $\gamma^{\text{vprDis}}_j$ &\hspace{-0.5cm} Discharge efficiency of vaporization plant\\
        $\beta$ &\hspace{-0.5cm} Boil-off gas coefficient\\
        $I^{\text{pipe}}_{\ell}$ &\hspace{-0.5cm} Initial capacity for pipeline $\ell$, [MMBtu/day]\\
        $U^{\text{pipe}}_{\ell}$ &\hspace{-0.5cm} Upper bound capacity for\\
        & pipeline $\ell$, [MMBtu/day]\\
        $\mathcal{I}^{\text{pipe}}_{\ell}$ &\hspace{-0.5cm} 1, if the pipeline $\ell$ exists; 0, otherwise\\
        $I^{\text{gStr}}_{j}$ &\hspace{-0.5cm} Initial storage capacity, [MMBtu]\\
        $I^{\text{vpr}}_{j}$ &\hspace{-0.5cm} Initial vaporization capacity, [MMBtu/d]\\
        $I^{\text{liq}}_{j}$ &\hspace{-0.5cm} Initial liquefaction capacity, [MMBtu/d]\\
        $I^{\text{store}}_{kj}$ &\hspace{-0.5cm} Initial capacity of storage facility\\
        $U^{\text{g}}_{\text{emis}}$ &\hspace{-0.5cm} Baseline emission of CO$_2$ in 1990\\
        &from non-generation consumption, [ton]\\
        $\Omega_n$ &\hspace{-0.5cm} representative day for day $n$\\
&\\
\midrule 
         \multicolumn{2}{l}{{\textbf{Investment Decision Variables}}}\\
      \midrule
         $x^{\text{op}}_{ni}\in \mathbb{Z}^+$ &\hspace{-0.5cm} Number of available plants\\         
         $x^{\text{est}}_{ni}\in \mathbb{Z}^+$ &\hspace{-0.5cm} Number of new plants established \\
         $x^{\text{dec}}_{ni}\in \mathbb{Z}^+$ &\hspace{-0.5cm} Number decommissioned plants \\
         $y^{\text{eCD}}_{nr}\in \mathbb{R}^+$&\hspace{-0.5cm} Charge/discharge capacity of storage battery\\
         $y^{\text{eLev}}_{nr}\in \mathbb{R}^+$&\hspace{-0.5cm} Battery storage level\\
         $z_{\ell}^{\text{e}}\in \mathbb{B}$&\hspace{-0.5cm} 1, if transmission line $\ell$ is built; 0, otherwise\\
         $z_{\ell}^{\text{g}}\in \mathbb{B}$&\hspace{-0.5cm} 1, if pipeline $\ell$ is built; 0, otherwise\\
&\\
\midrule 
 \multicolumn{2}{l}{{\textbf{Other Decision Variables for Power System}}}\\
      \midrule 
$p_{nti}\in \mathbb{R}^+$&\hspace{-0.4cm} Generation rate, [MW]\\
         $x_{nti}\in \mathbb{R}^+$ &\hspace{-0.4cm} Number of committed plants\\
         $x^{\text{down}}_{nti}\in \mathbb{R}^+$ &\hspace{-0.4cm} Number of plants shut-down\\
         $x^{\text{up}}_{nti}\in \mathbb{R}^+$ &\hspace{-0.4cm} Number of plants started up\\
         $f^{\text{e}}_{\ell t} \in {\mathbb{R}}$&\hspace{-0.4cm} Flow rates, [MW]\\
        $\theta_{nt}\in \mathbb{R}$ &\hspace{-0.4cm} Phase angle\\
        $s^{\text{eCh}}_{ntr},s^{\text{eDis}}_{ntr} \in \mathbb{R}^+$&\hspace{-0.4cm} Storage charged/discharged, [MW] \\
        $s^{\text{eLev}}_{ntr} \in \mathbb{R}^+$&\hspace{-0.4cm} Storage level, [MWh] \\
        $s^{\text{rem}}_{n\tau r} \in \mathbb{R}$&\hspace{-0.4cm} Storage carry over during\\
        & day $\tau$ for storage type $r\in \mathcal{S}^{\text{eL}} $\\
        $s^{\text{day}}_{n\tau r} \in \mathbb{R}^+$&\hspace{-0.4cm} Storage level at the beginning of\\
        &day $\tau$ for storage type $r\in \mathcal{S}^{\text{eL}}$\\
        $\kappa^{\text{capt}}_{nt} \in \mathbb{R}^+$&\hspace{-0.4cm} Captured CO$_2$ [ton/h] \\
        $\kappa^{\text{pipe}}_{n} \in \mathbb{R}^+$&\hspace{-0.4cm} CO$_2$ pipeline capacity [ton/h] \\
         $a^{\text{e}}_{nt}\in \mathbb{R}^+$ &\hspace{-0.4cm} Amount of load shedding, [MWh]\\
         $\mathcal{E}^{\text{e}}$ &\hspace{-0.4cm} Total emission from power system\\
&\\
\midrule 
 \multicolumn{2}{l}{\textbf{Other Decision Variables for }}\\
  \multicolumn{2}{l}{\textbf{NG System (all in MMBtu)}}\\
      \midrule  
         $x^{\text{gStr}}_{j}\in \mathbb{R}^+$ &\hspace{-0.4cm} Installed additional  storage capacities\\
          $x^{\text{vpr}}_{j}\in \mathbb{R}^+$ &\hspace{-0.4cm} Installed additional vaporization capacities\\
        $f^{\text{g}}_{\ell\tau} \in {\mathbb{R}}^+$&\hspace{-0.4cm} Flow between NG nodes \\
        $f^{\text{ge}}_{kn\tau} \in {\mathbb{R}^+}$&\hspace{-0.4cm} Flow from NG nodes to power nodes\\
        $f^{\text{gl}}_{kj\tau} \in {\mathbb{R}^+}$&\hspace{-0.4cm} Flow from node NG nodes to \\
        & \quad liquefaction plants\\
        $f^{\text{vg}}_{jk\tau} \in {\mathbb{R}^+}$&\hspace{-0.4cm} Flow rates from vaporization plants\\
        & \quad to NG nodes\\
         $g_{k\tau }\in \mathbb{R}^+$&\hspace{-0.4cm} NG supply (injection)\\
        $s^{\text{gStr}}_{j\tau}\in \mathbb{R}^+$&\hspace{-0.4cm} Storage capacities \\
        $s^{\text{vpr}}_{j\tau},s^{\text{liq}}_{j\tau} \in \mathbb{R}^+$&\hspace{-0.4cm} Vaporization and liquefaction amounts\\
         $a^{\text{g}}_{k\tau}\in \mathbb{R}^+$ &\hspace{-0.4cm} Amount of load shedding\\
        $a^{\text{LCDF}}_{k\tau}\in \mathbb{R}^+$ &\hspace{-0.4cm} Amount of LCDF consumption\\
        $\mathcal{E}^{\text{g}}$ &\hspace{-0.5cm} Total emission from NG system\\
        \bottomrule
\end{supertabular}

\section{Additional Results}

\subsection{Number of Representative Days}\label{app:rep-days}
\begin{figure}[htp]
    \centering
    \includegraphics[width=0.4\textwidth]{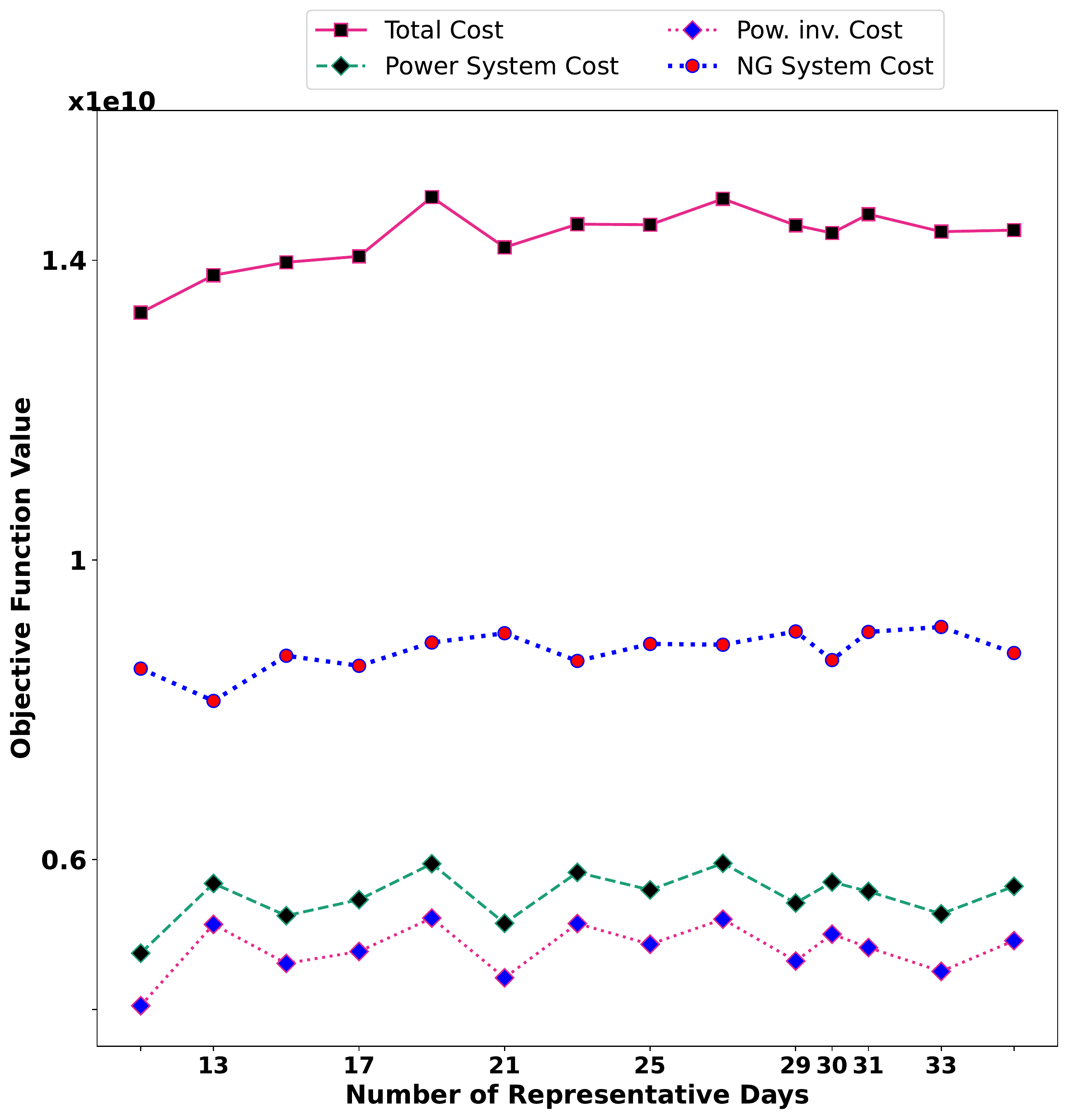}
    \caption{Scaled values for total system cost, power system cost, and NG system cost for different number of representative days}
    \label{fig:rep-costs}
\end{figure}

\begin{figure}[htp]
    \centering
    \includegraphics[width=0.5\textwidth]{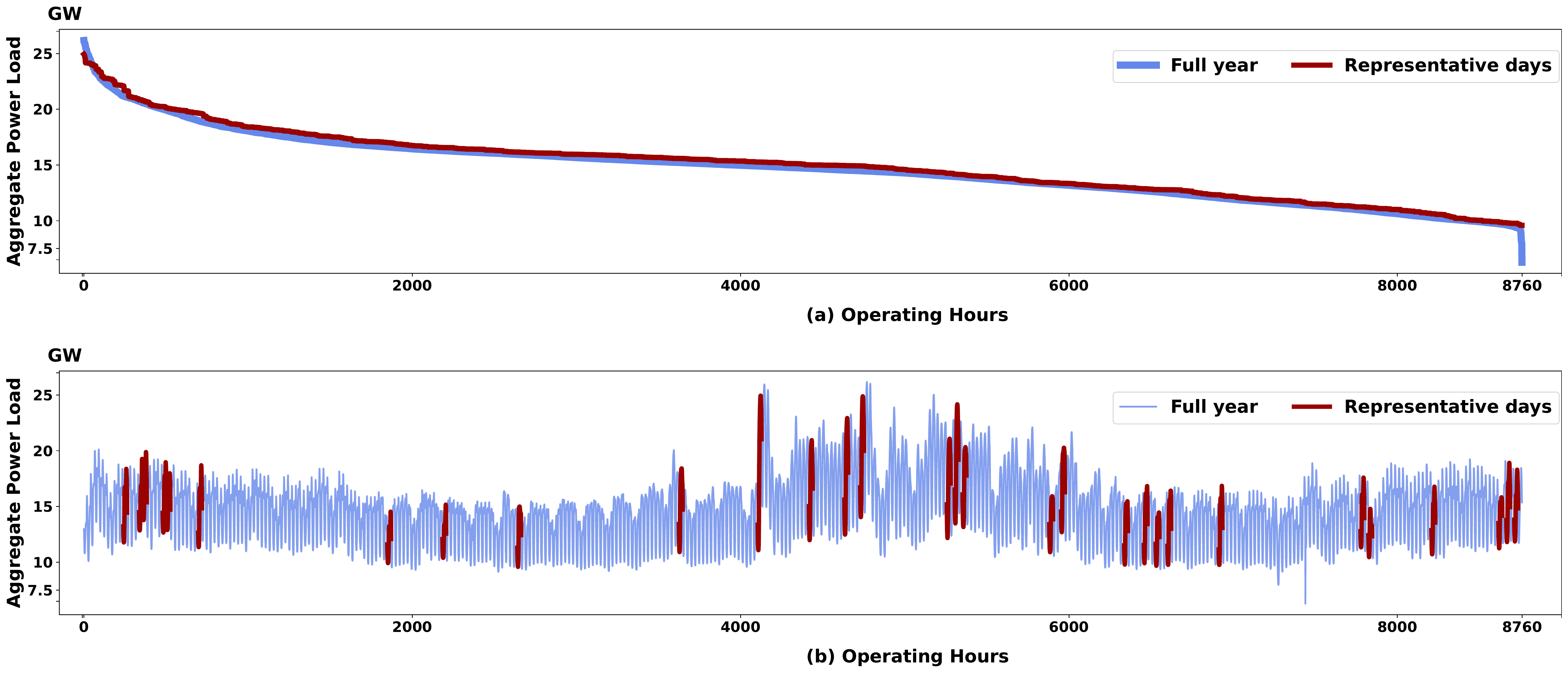}
    \caption{LDC (a) and actual load for the representative days and full year days.}
    \label{fig:LDC-rep-days}
\end{figure}

Fig.\ref{fig:rep-costs} shows the total system cost, power system cost, NG system costs as well as the investment cost in power system for different number of representative days. \bt{The cost values fluctuate for small number of representative days but tend to stabilize after 21 days. The total system cost exhibits little variation after 29 days, so we run all instances with 30 representative days. The electricity load duration curve (LDC) and the actual electricity demand of these 30 days are contrasted with the full-year data in Fig.~\ref{fig:LDC-rep-days}. We obtain the LDC by first assuming that the time series of electricity demand for the cluster medoid applies for all days in the cluster. We then sort the resulting time series in descending order.
Here, we see that the representative days, which cover different periods of the year, result in a reasonable approximation of the annual electricity LDC. Moreover, Fig.~\ref{fig:LDC-rep-days}b shows that the representative days account for some of the days with highest demand which is important for system planning purposes.}

\subsection{Generation Mix} \label{app:gen-mix}
\begin{figure*}
    \centering
    \includegraphics[width=0.95\linewidth]{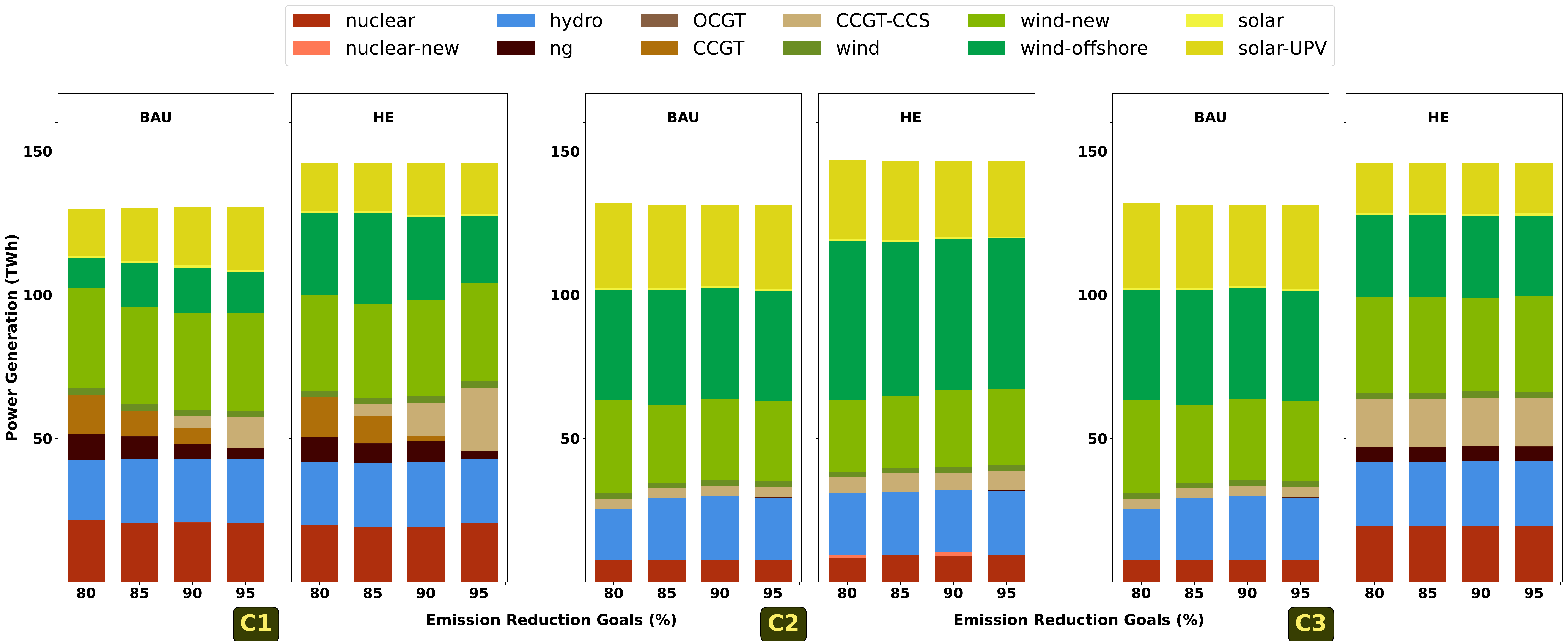}
    \caption{Generation mix of C1, C2, and C3}
    \label{fig:gen-mix1-2-3}
\end{figure*}

The generation mix of all cases for both scenarios is shown in Fig.~\ref{fig:gen-mix1-2-3}. 

\section{Problem Data}\label{App:input-data}
This section provides a description of the processes by which we obtained parameters for both networks for the New England Case study. We start with power system and explain the data preparation process. We then expound on steps we took to construct the NG network and associated data inputs.  In this study, our region is New England, but we want to emphasize that the modeling framework and the solution approach are applicable to other regions.




\subsection{Power Network Topology}
Our parameters for the power network are based on the \textit{US Test System} developed by \textit{Breakthrough Energy} \citep{PowerSimPaper2020}. The dataset contains high-temporal and spatial resolution load and variable renewable energy (VRE) data for the entire US ``base$\_$grid'' in the year 2016.
It also contains load data for the year 2020 and projections for the year 2030. The dataset provides a test system with detailed information for existing buses, substations, plants, and branches as well as generation profiles for existing renewable energy capacity, including solar and land-based wind.
We consider the base$\_$grid and start by filtering the data for New England region which corresponds to zone 1 to 6 in the dataset. We then filter for high voltage buses \citep{ReEDS2019} i.e. those with voltage greater than or equal to 345kV. This process results in 188 nodes each representing a high-voltage bus. The filtering process results in 192 transmission lines between them indicating that each node is connected to approximately one other node on average.  We use this information and the possible fact that some of the 188 nodes correspond to the same geographical location, in creating candidate transmission lines. Specifically, we assume that each candidate transmission line connects a node to its nearest one with a non-zero distance. We then estimate the susceptance and maximum flow of each candidate transmission line by a linear regression fitted on the parameters available for the existing lines.

Once the 188-node power network is constructed, we aggregate buses by their corresponding state and remove the transmission lines between buses of the same state. Aggregating nodes for zonal level is a common practice in the literature of capacity expansion for energy systems (see, for example, \cite{SepulvedaEtal2021, LiEtal2022, ZhaoEtal2017}). Reducing the number of nodes improves the scalability of the problem and increases the interpretability of the solution. For our case study, the aggregation process resulted in a network with 6 nodes (one node for each state in the New England region), and 63 transmission lines out of which 40 are candidate lines. \bt{Nevertheless, we only kept candidate transmission lines with the highest maximum flow between any two nodes. This process resulted in 7 candidate transmission lines.} Notice that there can be multiple existing transmission lines with varying characteristics (i.e. maximum flow, susceptance, and distance) between nodes.

\subsection{Power System Data}\label{app:power-data}

The ``base$\_$grid'' data contains power plant information at each bus. We first remove power plant types of ``dfo'' (distillate fuel oil) and ``coal'' as their share is not substantial in the base year and there are plans to completely phase out those plant types by 2050. We then calculate the existing generation capacity at each bus from each plant type by considering ``in-service'' plants with generation capacity greater than 10MW for ``ng'' (i.e., gas-fired plants) and ``nuclear'' plants, and greater than 2MW for VREs. Note that ``ng'' plants are treated as a lump since the Breakthrough Energy data set does not provide a breakdown between the capacity of combined cycle and open-cycle plants. We then assign each plant to the nearest node and eventually aggregate the generation capacity of each plant type for each of the 6 nodes. Table~\ref{tab:exis_plant_params} and \ref{tab:new_plant_params} present the technical assumptions for the existing and new power plants used in this study. Most of the parameters for the existing plants are derived from the Breakthrough Energy data set \cite{PowerSimData}. The footnote text in Table~\ref{tab:exis_plant_params} presents details on the value of each parameter. The technical assumptions for new plants are largely derived from National Renewable Energy Laboratories (NREL) Annual
Technology Baseline 2021 edition (ATB 2021) for the year 2045 \cite{ATB2021}. Values of parameters that are not available in ATB 2021 are adopted from the corresponding existing power plants or obtained from Sepulveda et al. \cite{SepulvedaEtal2021}. The details of each parameter is provided in the footnote text in Table~\ref{tab:new_plant_params}.


We incorporate regional capital cost multipliers provided in \textit{ReEDS Model Documentation} \cite{ReEDS2019}, summarized in Table~\ref{tab:regional-mults}, to distinguish between the capital cost of new power plants in the different model regions. These multipliers are applied to the baseline capital costs reported in Table~\ref{tab:new_plant_params}. Subsequently, the capital costs are annualized to be included in the single-stage investment planning model using the following formula for the annual cost fraction: $ \frac{\omega}{1-(\frac{1}{1+\omega})^{{lt}}}$. Here, $lt$ is the lifetime of the specific technology and $\omega$ corresponds to the discount rate of 7.1\%. Thus, the annualized CAPEX for new power plants is obtained by multiplying the  CAPEX by the annual cost fraction and regional multiplier. For every other investment cost (i.e. transmission lines, pipelines, storage etc.) we only multiply the CAPEX by the annual fraction factor to get the annualized CAPEX.

To obtain capacity factors (CFs) for existing renewables (solar, wind, and hydro) capacity, we used the hourly generation data available in the Breakthrough Energy data set, specifically, the \textit{base$\_$line} data \cite{PowerSimData}. We obtain the capacity factors of solar and wind plants by dividing their hourly generation by their nameplate capacity. The CFs at each node are obtained by taking the average over the CFs of the existing solar and wind plants belonging to that node. For the new solar and wind plants, we use the GenX data set which is obtained from the Future of Storage Study modeling of the U.S. Northeast \cite{GenX2017, MITEI2022FES}. The data set provides solar and wind capacity factors for 6 zones of the US Northeast in 2045 for 7 weather conditions corresponding to years 2007 to 2013. For each zone, the dataset contains 3 bins of wind and 2 bins of solar CF profiles reflecting the variation in VRE sites across each zone. Among the 6 zones used in the Future of Storage study modeling of the U.S. Northeast \cite{MITEI2022FES}, zone 1, 2, and 3 covers the New England region such that zone 1 represents Maine (ME), zone 2 corresponds to Massachusetts (MA), Vermont (VT), New Hampshire (NH) and Rhode Island (RI), and zone 3 represent Connecticut (CT). To use the Future of Storage study data for our 6-zone representation of the New England region, we made the assumption that the VRE capacity factors for new plants for MA, VT, NH, and RI for technology (solar, onshore wind) are the same. Further, we only allowed for offshore wind deployment in MA and CT.



\begin{figure}
    \centering
    \includegraphics[width=0.3\textwidth]{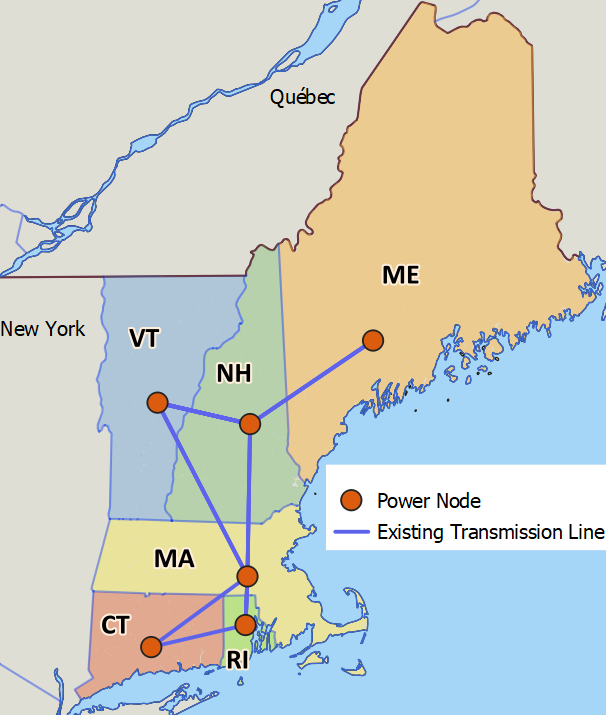}
    \caption{\bt{Node locations and existing connections between power nodes. Each connecting line can represent multiple transmission lines. Candidate transmission lines are not shown here.}}
    \label{fig:pow_top}
\end{figure}

Energy Storage is likely to be an essential part of the future power systems dominated by VRE supply. While many storage technologies are proposed or are under development, we only model the potential deployment of Li-ion battery type, given its commercial availability and available cost projections. We sourced Li-ion battery  cost and performance assumptions from \cite{ATB2021}, as summarized in Table~\ref{tab:storage-params}. Other economic and technical assumptions for the power network are presented in Table~\ref{tab:e-other-params} and Table~\ref{tab:res-avail}. Figure~\ref{fig:pow_top} illustrates a simplified power network.

\noindent\textbf{General Note about Tables:} The column of tables shows the associated symbol in the formulation. The notations with \textit{tilde} are \textit{crude} cost values whose manipulated forms (e.g. regional update, annualization) are used in the numerical model. For example,  $\tilde{C}^{\text{inv}}_i$ in Table~\ref{tab:new_plant_params} denote the value of ${C}^{\text{inv}}_i$ before annualization.

\begin{table}[ht]
\centering
\begin{tabular}{l|rrrrr}
\toprule
   & ng   & solar & wind & hydro & nuclear \\
   \midrule
CT & 4375 & 12    & 5    &       & 1888    \\
MA & 1763 & 3     & 40   & 0     & 0       \\
ME & 0    & 0     & 629  & 529   & 0       \\
NH & 1808 & 0     & 78   & 349   & 1226    \\
RI & 6667 & 369   & 65   & 1432  & 617     \\
VT & 0    & 61    & 111  & 199   & 0\\
\bottomrule
\end{tabular}
\caption{Capacity of existing plants at each state (MW)}
\end{table}

\begin{table*}[ht]
\caption{Parameter Values for Existing Plants}
    \centering
    \begin{tabular}{ll|rrrrrrrr}
    \toprule
      Symbol &  Parameter$\backslash$Type                     & ng     & solar    & wind   & hydro   & nuclear  \\
\midrule       
& FOM [\$/kW/year]$^1$             & 21     & 23       & 43         & 78       & 145      \\
$C^{\text{var}}_i$& VOM [\$/MWh]$^2$                 & 5      & 0        & 0          & 0        & 2        \\
$\eta_i$& CO$_2$ Capture [$\%$]$^3$        & 0      & -        & -          & -      & -   \\
$h_i$ & Heat Rate    [MMBtu/MWh]$^4$     & 8.7    & 0        & 0          & 0   & 10.6        \\
$\tilde{C}^{\text{dec}}_i$ & Decom. cost per plant [$\$$]$^5$ & 5.0e6  & 4.5e4    & 1e6        & -            & 3.0e8 \\
$U^{\text{prod}}_i$ & Nameplate capacity [MW]$^6$      & 173    & 6.3      & 42         & 23      & 933  \\
$L^{\text{prod}}_i$ & Min stable output [$\%$]$^7$       & 31     & 0        & 0          & 0       & 42      \\
$U^{\text{ramp}}_i$ & Hourly Ramp rate [$\%$]$^8$      & 96     & -        & -          & -        & 25      \\
$C^{\text{fix}}_i$ & FOM per plant   [\$/MW/year]$^{9}$ &3.6e6&1.45e5     & 1.8e6      & 1.8e6  & 1.4e8 \\
$C^{\text{startUp}}_i$ & Startup Cost [\$]$^{11}$& 4.52e4 & - & - & - & 4.6e4\\
\bottomrule
    \end{tabular}
    \label{tab:exis_plant_params}
    
    \footnotesize{$^{1}$ and $^{2}$ from \citep{ATB2021} in year 2019,\qquad $^3$ from \citep{eiaEmisRate2021},\qquad $^4$  approximated from the linear and quadratic coefficient of heat rate curve provided in \cite{PowerSimData},\qquad $^5$ estimated from \cite{Raimi2017} except for nuclear which is obtained from \citep{NukeDecom}. Decommissioning of hydro plants is not considered,\qquad $^6$ and $^7$  from \cite{PowerSimData},\qquad $^8$ estimated from 30-min ramp rate in \citep{PowerSimData},\qquad $^{9}$ $r1*1000*r6$ where $r1$ is the FOM value in $r1$ and $r6$ is nameplate capacity provide in row$6$,\qquad $^{11}$ from \cite{SepulvedaEtal2021}. The startup cost for  ``ng'' type is assumed to be the same as ``CCGT'' type.
    }
\end{table*}

\begin{table*}[ht]
\caption{Parameter Values for New Plants}
    \centering
    \scriptsize
    \begin{tabular}{ll|rrrrrrr}    
    \toprule
    
  Symbol &  Parameter$\backslash$Type                        &OCGT       & CCGT       & CCGT-CCS   & solar-UPV & wind-new     & wind-offshore & nuclear-new  \\
\midrule
&CAPEX [\$/kW]$^1$       & 780      & 935      & 2167     & 672       & 808      & 2043          & 6152     \\
&FOM [\$/kW/year]$^2$            & 21       & 27       & 65       & 15        & 35       & 74            & 145      \\
$C^{\text{var}}_i$& VOM [\$/MWh]$^3$               & 5        & 2        & 6        & 0         & 0        & 0             & 2        \\
$\eta_i$ &CO$_2$ Capture [$\%$]$^4$   & 0      & 0    & 90     & -         & -        & -             & -        \\
$h_i$&Heat Rate    [MMBtu/MWh]$^5$   & 9.72      & 6.36      & 7.16    & 0         & 0        & 0             & 10.46     \\
&Lifetime [year]          & 30       & 30       & 30       & 30        & 30       & 30            & 30       \\
$U^{\text{prod}}_i$& Nameplate capacity [MW]$^6$    & 237      & 573      & 400      & 10         & 10       & 10           & 360      \\
$\tilde{C}^{\text{inv}}_i$& CAPEX per Plant [$\$$]$^7$ & 1.85e8 &5.36e8 &8.67e8 & 6.72e6 & 8.01e6 &2.04e7 & 2.21e9 \\
$L^{\text{prod}}_i$& Minimum stable output [$\%$]$^8$ & 25       & 33       & 50       & 0         & 0        & 0             & 50      \\
$U^{\text{ramp}}_i$& Hourly ramping rate [$\%$]$^{9}$               & 100      & 100      & 100      & -       & -      & -           & 25       \\
$C^{\text{fix}}_i$& FOM per plant [\$/yr]$^{10}$ & 5.0e6 & 1.55e7 & 2.6e7 & 1.5e5  & 3.5e5 & 5.22e7      & 5.22e7 \\
$C^{\text{startUp}}_i$& Startup Cost [$\$$]$^{11}$       & 8.0e3 & 4.52e4 & 3.79e4 &- &-&- & 4.6e4       \\
\bottomrule
    \end{tabular}
    \label{tab:new_plant_params}
    \footnotesize{\\$^{1-5}$ from \citep{ATB2021} in year 2045. For CCGT-CCS, the ``Conservative'' technology class is considered. For ``wind-new'' and ``wind-offshore'',  ``Moderate-Class4'' technology class is considered. For all others ``Moderate'' cost assumption is considered,\qquad $^6$ from \citep{SepulvedaEtal2021} for ``OCGT'', ``CCGT'', CCGT-CCS, and ``nuclear-new''. For VRE, a modular capacity of 10MW is considered, \qquad $^7$ $r1*1000*r6$ where $r1$ is the CAPEX value in row$^1$ and $r6$ is nameplate capacity provide in row$^6$, \qquad $^8$ from \cite{SepulvedaEtal2021}, \qquad $^{9}$ from \citep{SepulvedaEtal2021}, \qquad $^{10}$ $r2*1000*r6$ where $r2$ is the FOM value in $r1$ and $r6$ is nameplate value provide in $r7$,\qquad $^{11}$ from \cite{SepulvedaEtal2021}
    }
\end{table*}

\begin{table*}[ht]
    \centering
        \caption{power storage parameters}
    \label{tab:storage-params}
    \begin{tabular}{ll|c c c}
         Symbol &  Parameter & Li-ion & Metal-air (low)$^1$ &Metal-air (high)$^2$   \\
         \midrule
     $\tilde{C}^{\text{EnInv}}$&    Energy capital cost [$\$$/kW]& 129$^3$ &0.1 &3.6\\
       $\tilde{C}^{\text{pInv}}$&  Energy power cost [$\$$ /kWh]& 156$^4$&595&950 \\
        $\gamma^{\text{eCh}}_r$&  Charge efficiency & 0.92$^5$ &0.7 &0.72\\
        $\gamma^{\text{eDis}}_r$&  Discharge efficiency & 0.92$^6$ &0.59 & 0.6 \\
       $C^{\text{EnInv}}_r$&   Energy related FOM ($\$$/kWh/year) & 3.22$^7$ &0 &100\\
        $C^{\text{pInv}}_r$&   Power related FOM ($\$$/kW/year) & 3.9$^8$ &14.9 &23.7\\
    $\gamma^{\text{selfD}}_{r}$ & hourly self-discharge rate & 2.08e-5$^9$ & 2.08e-5& 2.08e-5\\
         & Lifetime & 15$^{10}$ & 25 &25\\
         \bottomrule
    \end{tabular}
    
\footnotesize{columns $^1$ and $^2$ from \cite{MITEI2022FES},\\ $^{3}$ and $^4$ from \cite{ATB2021} in the year 2045 averaged over ``Advance'', ``Moderate'' and ``Conservative'' estimates, \qquad $^5$ and $^6$ from \cite{ATB2021} where the round-trip efficiency is provided at 85\%, \qquad  $^7$ and $^8$ 2.5$\%$ of energy capital and power cost (row 1 and 2), respectively  \cite{ColeFrazier2021}, \qquad $^9$ from \cite{MITEI2022FES} the monthly self-discharge provided at 1.5\%, \qquad $^8$ from \cite{ATB2021}}
\end{table*}

\begin{table*}[ht]
\caption{Other parameters for power network}
    \centering
    \begin{tabular}{ll|r}
    \toprule
     Symbol &  Parameter & Value\\
    \midrule 
    & Transmission line lifetime [year]&  30\\
    & Weighted average cost of capital (WACC)$^{1}$ & $7.1\%$\\
 $\tilde{C}^{\text{trans}}_\ell$&    Transmission line investment cost [$\$/$MW/mile]$^2$ &3500 \\
    &Li-ion Battery lifetime [year]$^3$ & 30\\
    &Transmission line lifetime [year]$^4$ & 30\\
    & Uranium price [$\$$/MMBtu]$^5$ & 0.72\\
    \end{tabular}
    \label{tab:e-other-params}
    
    \footnotesize{$^1$ from \cite{SepulvedaEtal2021}, \qquad  $^2$ from \cite{ReEDS2019},\qquad $^3$ from \cite{ATB2021}, \qquad  $^4$ from \citep{ATB2021} in year 2045, $^5$ from \citep{SepulvedaEtal2021} in year 2045
    }
\end{table*}

\begin{table}[ht]
\caption{Resource Availability Data}
    \centering
    \begin{tabular}{l|r}
    \toprule
     $\mathcal{Q}$ & Maximum Available Value\\
    \midrule 
    {``solar'', solar-UPV} & 22 GW\\
    {``wind'', ``wind-new''} & 10 GW\\
    {wind-offshore} & 280 GW\\
    {``nuclear'', ``nuclear-new''} & 3.5 GW\\
    \end{tabular}
    \label{tab:res-avail}
    
    \footnotesize{Resource availability amounts are obtained from \cite{E3Team2020} 
    }
\end{table}

\begin{table*}[ht]
\caption{Regional CAPEX multipliers for new plant types}
\label{tab:regional-mults}
\centering
\begin{tabular}{c| rrrrrrr}
\toprule
State/Technology &OCGT   & CCGT   & CCGT-CCS & solar-UPV & wind & wind-offshore & nuclear \\
\midrule 
Connecticut (CT)      & 1.25 & 1.3  & 1.3    & 1.15      & 1.4      & 1.1             & 1.1         \\
Massachusetts (MA) & 1.1  & 1.1  & 1.1    & 1.05      & 1.35     & 1.1             & 1.05        \\
Maine (ME) & 1.25 & 1.3  & 1.3    & 1.1       & 1.35     & 1.1             & 1.1         \\
New Hampshire (NH)   & 1.1  & 1.1  & 1.1    & 1.05      & 1.35     & 1.1             & 1.05        \\
Rhode Island (RI) & 1.2  & 1.25 & 1.25   & 1.1       & 1.35     & 1.1             & 1.05        \\
Vermont (VT) & 1.1  & 1.1  & 1.1    & 1.05      & 1.35     & 1.1             & 1.05 \\
\bottomrule
\end{tabular}
\end{table*}

\subsection{NG Network Topology and Data}\label{app-NG-topology}
NG network consists of two types of nodes: 1) NG nodes: each NG node has an injection capacity and load; 2) SVL nodes: each SVL node consists of Storage, Vaporization, and Liquefaction (SVL) facilities each with their own capacity. We construct the NG network based on the data available on Energy Information Administration (EIA) website \citep{eiaWebsite2021} as well as \citep{EIA-layer-maps}. We use the former reference to obtain information about inter-state pipelines. The inter-state pipeline data provide information on the counties connected through a uni-directional pipeline as well as the capacity of each connecting pipeline. We first filter all the pipelines that connect  counties inside the region or connect a county outside of New England to a county inside. We consider counties inside the New England region as the NG network nodes and pipelines between them as edges. We then consider pipelines connecting nodes to outside counties and calculate the maximum injection capacity (i.e. supply) for each node.

\bt{Once the inter-state pipelines are determined, we use the New England  NG pipeline data provided in \citep{EIA-layer-maps} to estimate the existing intra-state pipelines in the region. The information about intrastate pipelines is limited, so we assume that the capacity of these pipelines is the same as the injection capacity at their origin nodes. This process results in 18 nodes and {28  NG pipelines} to make up the existing NG network. Note that there is only at most one pipeline between each pair of nodes, however, the formulation is for the general case where, similar to the power system, there can be multiple pipelines between NG nodes.}


NG is usually stored in its liquefied form called LNG. LNG storage usually involves three types of facilities:
\begin{itemize}
    \item Storage tank: smaller tanks located inland that are filled by truck supply from the import facilities and are used to regulate the pressure of the gas system – these facilities have tanks and vaporization facilities \cite{MallapragadaEtal2018}. The capacity of a storage facility is measured in energy/volume level (i.e. MMBtu or MMScf)
    \item Liquefaction: a facility that receives NG from a pipeline  and liquefies it at a temperature around -162 Celsius \cite{MallapragadaEtal2018, Mesko1996}. The capacity of liquefaction facility is measured in terms of flow rate (i.e. MMBtu/period or MMScf/period)
    \item Vaporization: The facility evaporates the LNG leaving the storage facility by warming it with seawater or air to produce gas that is injected into the pipeline network. In certain areas, trucks are used to transport LNG from a major storage site to a smaller storage site where they are later evaporated \cite{MallapragadaEtal2018}. For ease of transportation, storage, and vaporization facilities are built at the same location. The capacity of vaporization facility is measured in terms of flow rate (i.e. MMBtu/day or MMScf/day).
\end{itemize}

\begin{figure}
    \centering
    \includegraphics[width=0.4\textwidth]{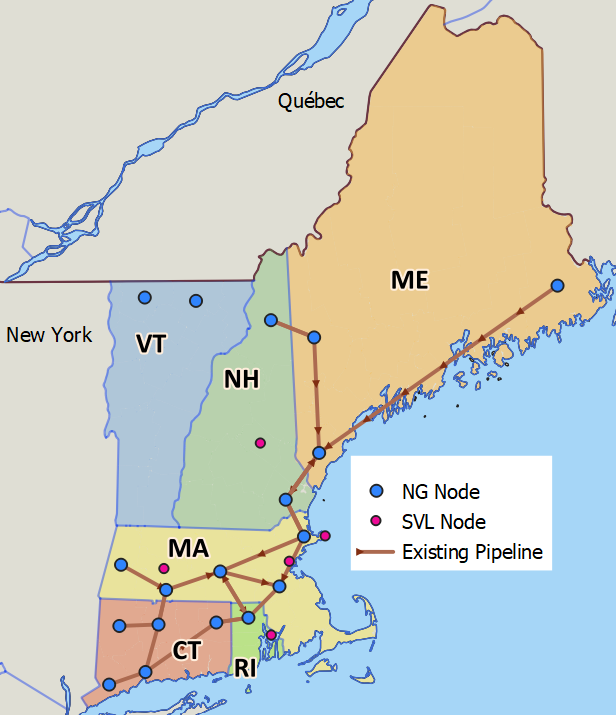}
    \caption{\bt{Node locations and existing connections between NG
nodes. Each connecting line can represent multiple pipelines. Candidate pipelines as well as the connections between SVL and NG nodes are not shown here}}
    \label{fig:ng_top}
\end{figure}

The simplified topology of the NG system is depicted in Figure~\ref{fig:ng_top}.

We construct the SVL network based on the data provided in \cite{NGstatGuide2021}. We assume that all three facilities are located at the same location. There are currently 5 liquefaction facilities and 43 storage in New England. The exact location of these facilities is not provided, but the source provides a map of the region showing the approximate locations of storage tanks. Given that there are only 5 liquefaction facilities, we first cluster storage tanks into 5 locations and assume a liquefaction and a vaporization facility at each location. This assumption effectively approximates the practice of moving LNG via trucks from the centralized liquefaction facilities to the distributed storage (and vaporization) facilities. Each of these locations is a node in the SVL network. The total liquefaction, vaporization, and storage capacities for the New England region are given in \cite{NGstatGuide2021}. To account for variation in the capacities, we unevenly distribute these capacities over 5 SVL nodes shown in Figure \ref{fig:ng_top}. Moreover, there are two LNG import locations in New England, each having storage and vaporization facilities with given capacities. We consider these two locations as SVL nodes with no liquefaction capacity,  resulting in 7 SVL nodes in total. The transportation of NG between different nodes is assumed to be realized as follows:

\begin{itemize}
    \item NG node to NG node: We assume that there is a candidate pipeline between each node and \bt{two of its closest nodes, resulting in 36 candidate pipelines}. The capacity of candidate pipelines is set to the average capacity of the existing pipelines. 
    \item NG to power node: The number of NG nodes is three times more than power nodes (6 nodes in power system vs. 18 nodes in NG system). We use this fact to assume that each power node is already connected to  three nearest NG nodes via distribution pipelines. The pipeline capacity available to these neighboring NG nodes thus limits the amount of flow between NG and power nodes, so we assume that the connection between NG and power nodes has no capacity limits.
    \item NG and SVL nodes: We assume that each NG node is already connected to two nearest SVL nodes through two sets of pipelines. The first set carries NG from NG node to the liquefaction facility in an SVL, and the second set of pipelines transports NG from a vaporization facility in an SVL node to an NG node. 
\end{itemize}

\begin{figure}
    \centering
    \includegraphics[width=0.35\textwidth]{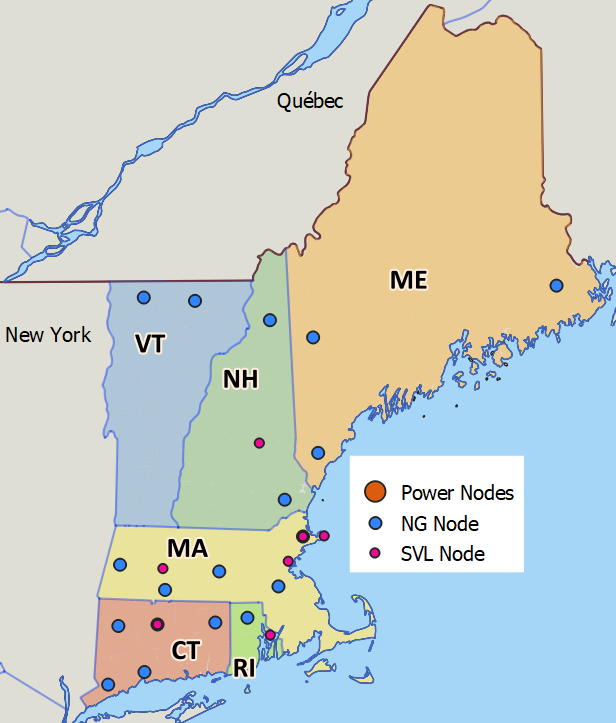}
    \caption{\bt{Three types of nodes in the formulation including power, NG, and storage-vaporization-liquefaction (SVL) nodes. Note that two pairs of SVL and NG nodes share the same location. The connections are not shown here.}}
    \label{fig:all_nodes}
\end{figure}

Figure~\ref{fig:node-connections} shows our assumption about connection between different nodes. The parameters for NG system are presented in Table~\ref{tab:ng-other-params}. Figure~\ref{fig:all_nodes} shows the dispersion of power, NG, and SVL nodes in the New England region.

\begin{table*}
\caption{Other parameters for NG/SVL network}
    \centering
    \begin{tabular}{ll|r}
    \toprule
    Parameter & Value\\
    \midrule 
  $\tilde{C}^{\text{strInv}}_j$&  Storage tank CAPEX$^1$ [$\$$/MMBtu]& 729.1\\
    $\tilde{C}^{\text{vprInv}}_j$&  Vaporization CAPEX$^2$ [$\$$/MMBtu] & 1818.31\\
    $C^{\text{strFix}}_j$&  Storage tank FOM$^3$ [$\$$/MMBtu]& 3.6\\
   $C^{\text{vprFix}}_j$& Vaporization FOM$^4$ [$\$$/MMBtu/d] & 327.3\\
   $\gamma^{\text{ligCh}}_j$ & Liquefaction charge efficiency ($\%$) & 100\\
   $\gamma^{\text{vprDis}}_j$ & Vaporization discharge efficiency$^5$ ($\%$) & 98.9\\
  &  Pipeline lifetime [year] & 30\\
 $C^{\text{gShed}}_j$  & NG load shedding cost [$\$/$MMBtu] & 1000\\
  $\eta^{\text{g}}$&  Emission factor for NG $^{6}$ [ton/MMBtu]& 0.053 \\
 $C^{\text{alt}}$ & LCDF price [\$/MMBtu]$^{7}$ &20\\
  & SVL lifetime [year] & 30\\
    $\tilde{C}^{\text{pipe}}_\ell$& Pipeline investment cost [$\$$/mile]$^{8}$ & 5.34e6\\
   $C^{\text{ng}}$& NG price [$\$/$MMBtu]$^{9}$  & 5.45 \\
   \bottomrule
    \end{tabular}
    \label{tab:ng-other-params}
    
    \footnotesize{$^1$ from Table 1 of \cite{LNGcapex2019},\qquad $^2$ from Table 1 of \cite{LNGcapex2019},\qquad $^3$ 0.5$\%$ of CAPEX according to \cite{Mesko1996},\qquad $^4$ 1.8$\%$ of CAPEX according to \cite{Mesko1996},\qquad $^5$ estimated from \cite{MallapragadaEtal2018},
    \qquad $^{6}$ from \cite{eiaWebsite2021},
    \qquad $^{7}$ from \cite{ColeEtal2021},
    \qquad $^{8}$ approximated from \citep{PipePrice2021}, \qquad $^{9}$ from \cite{SepulvedaEtal2021}
    }
\end{table*}

\subsection{Electricity and NG Demand Scenarios} \label{app:electrification-scen}

We use NREL's Electrification Future Study Load Profile dataset \cite{EFSload2022} data for power system in both high and reference electrification scenarios. 

The load for NG is generally disaggregated into five sectors including residential, commercial, industrial, vehicle fuel consumption, and electric power customers \cite{eiaWebsite2021}. The consumption for electric power consumers is a decision variable in our model, so our input for NG demand involves the NG demand in the remaining four sectors. 
Our approximation of daily NG consumption profile for residential and commercial sectors under various electrification scenarios is based on NREL's ``End-Use Load Profiles for the U.S. Building Stock'' project \cite{ResstockComstock2022} and NREL's ``Electric Technology Adoption and Energy Consumption'' \cite{EFSETA2022}. The former project consists of ResStock and ComStock projects that are analysis tools to help users better understand the residential and commercial building stock energy use. Each project has a website that allows users to get state-level consumption of NG for a \textit{typical meteorological year (TMY)} with a 15-minute and daily resolution for various end-use subsectors such as residential and commercial space heating. The latter provides a projection of annual energy consumption for various subsectors (e.g., space heating) under different electrification scenarios between 2017 and 2050.

We first consider residential and commercial sectors and obtain their daily consumption trend from TMY profiles available from ReStock and ComStock \cite{ResstockComstock2022}. We then scale the values of each subsector by their corresponding annual consumption of NG in reference electrification scenario in 2050 using the data available at \cite{EFSETA2022}. For high electrification scenario, we repeat the same process for all subsectors except space heating for which we scale its values based on the annual space heating consumption in 2050 under the high electrification scenario.

The monthly state-level NG load for all five sectors is available in the EIA website \cite{eiaWebsite2021}. The EIA website does not provide any information on the distribution of the monthly demand over its days. Therefore, we consider industrial and vehicle fuel consumption in 2016 and uniformly distributed the monthly load across their corresponding days. We then scale the values based on the annual industrial and vehicle fuel consumption in 2050 under reference and high electrification scenarios. Finally, we aggregate loads for all sectors and subsectors to obtain NG consumption of each state in 2050 under the two electrification scenarios. 


Once the daily NG demand is obtained for each state, we disaggregate the demand over each county based on the population of counties in 2019 \citep{CensusData2019} as well as their industrialization level. We first obtain share of each county in the state's population. We then obtain the industrialization level of counties from data provided in the NREL report on county-level energy use \citep{NRELcountyEnergy2021} which estimates the usage of industrial energy (net power, NG, coal, etc.) for each sector (manufacturing, agriculture, construction, and mining). Using this dataset, we estimate the share of industrial NG consumption of each county in the state's total industrial consumption. Subsequently, we compute the demand of each county by multiplying its population share to the state's NG consumption in residential, commercial, and vehicle fuel consumption sectors, and multiplying its industrial share by the state's NG consumption in industrial sector. Finally, we assign each county's load to its nearest NG node.

\subsection{Emission Amounts}\label{app:emission-amount}

All England states have set a goal to reduce the emission of GHG by at least 80$\%$ until 2050  below the baseline years which is 1990 for all all states except Connecticut \cite{Brattle2019}. The total CO$_2$ emission for the New England states was 171.2 metric tons (mt) in 1990, of which 43.9mt was electricity energy-related emission and 23.6mt was NG energy-related emission \cite{eiaCO2emis}. The remaining emission was caused by consuming coal and petroleum that we do not consider in this study. Based on these figures, $U^{\text{g}}_{\text{emis}} =$ 23.6e6 and $U^{\text{e}}_{\text{emis}}=$43.9e6.

\begin{table}[htp]
\centering
\caption{Actual emission budget (ton)}
\begin{tabular}{l|llll}
\toprule
      & $\zeta=80\%$ & $\zeta=85\%$ & $\zeta=90\%$ & $\zeta=95\%$ \\
      \midrule
C1 & 3.51e7     & 3.73e7     & 3.95e7     & 4.17e7     \\
C2 and C3   & 5.4e7     & 5.74e7     & 6.08e7     & 6.41e7   \\
\bottomrule
\end{tabular}
\end{table}

\subsection{CCS parameters}
We based our estimation of CCS parameters on \cite{TeletzkeEtal2018, BlondesEtal2013, BoardEtal2019}. We assume that the collected carbon is stored in Appalachian Basin at a Marcellus region located in the middle of Pennsylvania \cite{BlondesEtal2013}. We then calculate the distance between each node and the storage site. The total capacity of the Appalachian Basin is 1278 Mt (megaton). Assuming the Basin is operable for 100 years, the total annual carbon storage capacity $U^{\text{CCS}}$ becomes 12.78 Mt. Other parameters are calculated as follows:
\begin{itemize}
    \item $C^{\text{inv}}_{\text{CO2}}$: Reference \cite{BoardEtal2019} provides CAPEX for 10 and 100 miles pipelines. We consider 100-mile pipelines as all distance values are greater than 200 miles. The CAPEX and FOM for 100 miles pipeline are 225 $\$$M and 1.3 $\$$M (million dollars), respectively. With 30 years of lifetime for CO$_2$ pipelines and WACC=7.1\%, the CAPEX becomes 18.31 $\$$M. The FOM is given at 1.3 $\$$M, so the per-mile investment and FOM is (18.31e6+1.3e6)/100 = 196e3 $\$$/mile. The reference assumes that the capacity of pipeline is 10 Mt/y (megaton per year). Therefore, the levelized investment and FOM becomes 196e3/10e6 = 0.0196 $\$$/mile/ton.
    \item $C^{\text{str}}_{\text{CO2}}$: The Appalachian basin is an aquifer type storage. The CAPEX is given at 4.3 $\$$M in \cite{BoardEtal2019} for a storage site of type aquifer with 7.3 Mt capacity per year. The CAPEX consists of injection site screening and evaluation, injection equipment, and drilling 6 wells. Assuming 30 years of lifetime, the annualized CAPEX becomes 350e3 $\$$. The FOM is given at 600e3 $\$$, so the levelized investment and FOM is (350e3+600e3)/7.3e6 = 0.13 $\$$/ton. 
    \item $E^{\text{pipe}}$: For a pipeline of 100 miles long with capacity 10 Mt/year, reference \cite{BoardEtal2019} gives the electric requirement at 32,000 MWh/year or equivalently 32/8760e6 = 0.00365e-6 [MWh/mile/ton/hour].
    \item $E^{\text{pipe}}$: For a pipeline with capacity 10 Mt/year, each pump consumes 4190 MWh electricity annually which accounts for 4190/(8760e6)=0.478e-6 MWh/ton/hour. 
    \item $E^{\text{cprs}}$: Compression pump are located every 3.3 miles along the pipeline \cite{BlondesEtal2013}. Therefore, the value is $d_n$/3.3.
\end{itemize}

We assume that CAPEX for compression pumps is negligible. Also note that these cost estimations for CCS storage are conservative as we levelized the investment and FOM costs, and do not consider other cost parameters such as labor, FOM of compression pumps and fugitive emission amount which are listed in \cite{BlondesEtal2013}.

\bibliographystyle{plain}
\bibliography{AE-Arxiv.bib}

\end{document}